\documentclass[a4paper,11pt]{article}
\pdfoutput=1 % if your are submitting a pdflatex (i.e. if you have
\usepackage{jheppub} % for details on the use of the package, please
\usepackage[T1]{fontenc} % if needed

\usepackage[T1]{fontenc} % Works for Umlaute
\usepackage[utf8]{inputenc} % Input
\usepackage{microtype} % Subliminal refinements towards typographical perfection
\usepackage{lmodern} % Better Font
\usepackage{booktabs} % Nicer tables
\usepackage{mdframed} % Framed boxes \begin{mdframed}
\usepackage{graphicx}

\usepackage{hyperref}
\usepackage{xcolor}
\definecolor{dark-red}{rgb}{0.4,0.15,0.15}
\definecolor{dark-blue}{rgb}{0.15,0.15,0.4}
\definecolor{medium-blue}{rgb}{0,0,0.5}

\hypersetup{%
	pdftitle   = {Massive carrollian fields at timelike infinity},
	pdfkeywords = {Carrollian geometry, Wigner method, representation theory, celestial holography},
	pdfauthor  = {Emil Have, Kevin Nguyen, Stefan Prohazka, Jakob Salzer},
	colorlinks, linkcolor={dark-red}, citecolor={dark-blue}, urlcolor={medium-blue}
}

\usepackage{amsmath,amsfonts,amssymb}
\usepackage{mleftright} \mleftright
\usepackage{tensor} % Tensor with \indices{^ab_cd}
\usepackage{braket} % Braket works \bra \ket \braket{ | }
\usepackage{bm} % Bold math
\usepackage{mathtools} % Useful toosl for AMS Math

\usepackage{mathrsfs} % Nice fonts for Lagrangian, Hamiltonian, Path intgralmeasure \mathscr{LHD}

\providecommand*{\dd}{\mathop{}\!d}
\renewcommand*{\dd}{\mathop{}\!d}

\providecommand*{\pd}{\partial}
\renewcommand*{\pd}{\partial}

\providecommand*{\cd}{\mathop{}\!\nabla}
\renewcommand*{\cd}{\mathop{}\!\nabla}

\providecommand*{\Ld}{\mathop{}\!\mathsterling} % Lie derivative
\renewcommand*{\Ld}{\mathop{}\!\mathsterling}

\providecommand*{\RR}{{\mathbb{R}}}
\renewcommand*{\RR}{{\mathbb{R}}}

\newcommand{\zAdSC}{\mathsf{AdSC}}

\newcommand{\zAdS}{\mathsf{AdS}}

\newcommand{\zTi}{\mathsf{Ti}}
\newcommand{\zSpi}{\mathsf{Spi}}

\newcommand{\scri}{\mathscr{I}}

\newcommand{\g}{\mathfrak{g}}
\newcommand{\m}{\mathfrak{m}}

\newcommand{\h}{\mathfrak{h}}

\renewcommand{\d}{\partial}

\newcommand{\so}{\mathfrak{so}}
\newcommand{\iso}{\mathfrak{iso}}

\newcommand{\ISO}{\operatorname{ISO}}
\newcommand{\SO}{\operatorname{SO}}

\newcommand{\x}{\bm{x}}
\newcommand{\y}{\bm{y}}

\renewcommand{\k}{\bm{k}}

\newcommand{\bB}{\bm{B}}
\newcommand{\HH}{\mathbb{H}}

\newcommand{\MM}{\mathbb{M}}

\newcommand{\ad}{\operatorname{ad}}

\newcommand{\hphi}{\hat{\phi}}
\newcommand{\bzero}{\boldsymbol{0}}
\newcommand{\D}{{\partial}}
\usepackage{tikz,pgfplots}

\usetikzlibrary{shapes.misc,shapes.symbols,snakes,decorations.text}
\usetikzlibrary{math}
\usetikzlibrary{calc}
\usetikzlibrary{positioning}
\tikzset{cross/.style={cross out, draw=black, thick, minimum size=2*(#1-\pgflinewidth), inner sep=0pt, outer sep=0pt},
	cross/.default={3pt}}
\pgfdeclarelayer{nodelayer}
\pgfdeclarelayer{edgelayer}
\pgfsetlayers{edgelayer,nodelayer,main}
\tikzstyle{ghost}=[fill=none, draw=none, shape=circle]
\tikzstyle{dot}=[fill=black, draw=black, shape=circle, scale=0.5]
\tikzstyle{grey line}=[-, draw={rgb,255: red,128; green,128; blue,128}]
\tikzstyle{blue line}=[-, draw=blue,thick]%, fill={rgb,255: red,162; green,246; blue,255}]
\tikzstyle{red line}=[-, draw=red,thick]%, fill={rgb,255: red,255; green,100; blue,100}]
\tikzstyle{blue line2}=[-, draw=blue]
\tikzstyle{dash blue}=[-, draw=blue, dashed]
\tikzstyle{dash red}=[-, draw=red, dashed, thick]
\tikzstyle{dash black}=[-, draw=black, dashed, thick]
\tikzstyle{dash grey}=[-, draw={rgb,255: red,128; green,128; blue,128}, dashed]
\tikzstyle{thick black line}=[-, thick]
\tikzstyle{thick black line dashed}=[-, thick,dashed]
\tikzstyle{blue fill}=[-, draw=none, fill={rgb,255: red,126; green,214; blue,255}]
\tikzstyle{black line}=[-]
\tikzstyle{thick red}=[-,draw=red,thick]
\tikzstyle{red fill}=[-, fill={rgb,255: red,255; green,162; blue,164}, draw={rgb,255: red,255; green,0; blue,4}]
\tikzstyle{purple fill}=[-, fill={rgb,255: red,128; green,0; blue,255}, draw={rgb,255: red,100; green,15; blue,128}]
\tikzstyle{arrow}=[draw=black, <-]
\tikzstyle{green arrow}=[draw=ForestGreen, ->]
\definecolor{colour1}{RGB}{252,57,0}
\definecolor{colour2}{RGB}{252,115,0}
\definecolor{colour3}{RGB}{252,173,0}
\definecolor{colour4}{RGB}{252,202,0}
\definecolor{colour5}{RGB}{252,255,130}
\definecolor{contournoin}{RGB}{255,255,0}
\definecolor{contournoout}{RGB}{228,0,5}

\title{\boldmath Massive carrollian fields at timelike infinity}

\author[a,1]{Emil Have\note{\href{https://orcid.org/0000-0001-8695-3838}{ORCID: 0000-0001-8695-3838}},}
\author[b,2]{Kevin Nguyen,\note{\href{https://orcid.org/0000-0002-7714-2328}{ORCID: 0000-0002-7714-2328}}}
\author[c,3]{Stefan Prohazka\note{\href{https://orcid.org/0000-0002-3925-3983}{ORCID: 0000-0002-3925-3983}}}
\author[b,4]{and Jakob Salzer\note{\href{https://orcid.org/0000-0002-9560-344X}{ORCID: 0000-0002-9560-344X}}}

\affiliation[a]{Niels Bohr International Academy, Niels Bohr Institute,\\
	University of Copenhagen, Blegdamsvej 17, DK-2100 Copenhagen Ø, Denmark}
\affiliation[b]{Université Libre de Bruxelles and International Solvay Institutes,\\ ULB-Campus Plaine
	CP231, B-1050 Brussels, Belgium}
\affiliation[c]{University of Vienna, Faculty of Physics, Mathematical Physics,\\
	Boltzmanngasse 5, 1090, Vienna, Austria}

\emailAdd{emil.have@nbi.ku.dk}
\emailAdd{kevin.nguyen2@ulb.be}
\emailAdd{stefan.prohazka@univie.ac.at}
\emailAdd{jakob.salzer@ulb.be}

\abstract{Motivated by flat space holography, we demonstrate that massive spin-$s$ fields in Minkowski space near timelike infinity are massive carrollian fields on the carrollian counterpart of anti-de Sitter space called $\mathsf{Ti}$. Its isometries form the Poincaré group, and we construct the carrollian spin-$s$ fields using the method of induced representations. We provide a dictionary between massive carrollian fields on $\mathsf{Ti}$ and massive fields in Minkowski space, as well as to fields in the conformal primary basis used in celestial holography. We show that the symmetries of the carrollian structure naturally account for the BMS charges underlying the soft graviton theorem. Finally, we initiate a discussion of the correspondence between massive scattering amplitudes and carrollian correlation functions on $\mathsf{Ti}$, and introduce physical definitions of detector operators using a suitable notion of conserved carrollian energy-momentum tensor.}

\preprint{UWThPh 2024-6}

\begin{document} 
	\maketitle
	\flushbottom
	
	\section{Introduction}
	\label{sec:introduction}
	
	More than two decades after the advent of the AdS/CFT correspondence
	\cite{Maldacena:1997re,Gubser:1998bc,Witten:1998qj}, substantial
	effort is currently devoted to the investigation of the holographic
	principle in more general settings. In particular the program of
	\textit{celestial holography} aims at providing a holographic
	description of quantum gravity in asymptotically flat spacetimes in
	terms of a somewhat exotic conformal field theory (CFT) defined on the
	two-dimensional celestial sphere $\mathbb{CS}^2$
	\cite{Pasterski:2021raf,McLoughlin:2022ljp,Donnay:2023mrd}. Within
	this perspective the Lorentz group is realised as the group of global
	conformal transformations of the celestial sphere $\mathbb{CS}^2$,
	while translations are realised as internal symmetries connecting
	conformal primaries of different scaling dimensions. Alternatively one
	can choose to realise the full BMS group
	\cite{Bondi:1962px,Sachs:1962zza,Sachs:1962wk} and its Poincaré
	subgroup as conformal isometries of the spacetime null conformal
	boundary $\scri \cong \mathbb{R} \times \mathbb{CS}^2$, yielding a
	variant of flat holography called \textit{carrollian holography}. This
	has allowed in particular to interpret massless scattering amplitudes
	as a set of correlators of a three-dimensional conformal carrollian
	theory living on $\scri$
	\cite{Bagchi:2016bcd,Banerjee:2018gce,Banerjee:2019prz,Donnay:2022aba,Bagchi:2022emh,Donnay:2022wvx,Bagchi:2023fbj,Saha:2023hsl,Salzer:2023jqv,Nguyen:2023vfz,Saha:2023abr,Nguyen:2023miw,Bagchi:2023cen,Mason:2023mti,Chen:2023naw}.
	Although this carrollian approach has attractive features, only
	massless particles have been accounted for thus far, and the purpose
	of the present work is to provide a carrollian description of massive
	particles.
	
\begin{figure}[h!]
		\centering
		\includegraphics[width=0.99\textwidth]{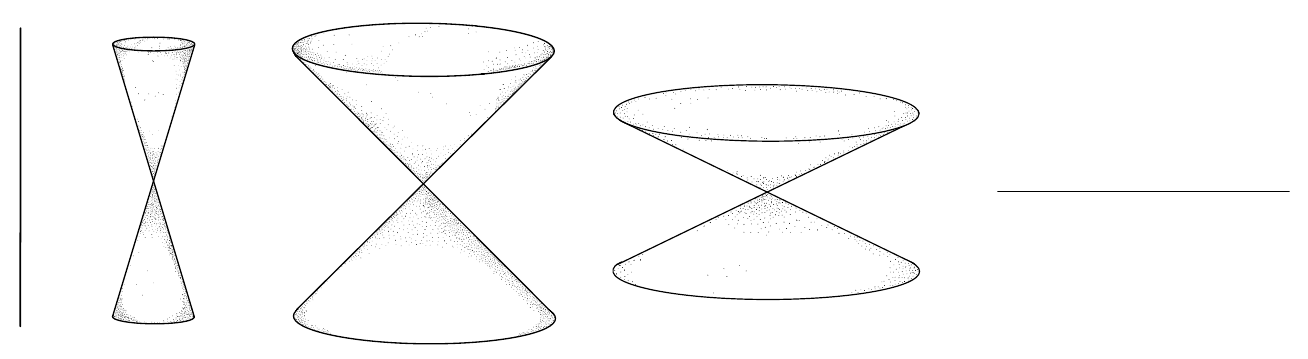}
		\begin{tikzpicture}[overlay]
		\begin{pgfonlayer}{nodelayer}
		\node [style=ghost] (0) at (-14.8, -0.5) {{$c=0$}};
		\node [style=ghost] (0) at (-13.3, -0.5) {{$c\ll 1$}};
		\node [style=ghost] (0) at (-10.2, -0.5) {{$c=1$}};
		\node [style=ghost] (0) at (-6.2, -0.5) {{$c\gg 1$}};
		\node [style=ghost] (0) at (-2, -0.5) {{$c=\infty$}};
		\end{pgfonlayer}
		\end{tikzpicture}
		\vspace{0.7cm}
		\caption{Lightcones opening and closing as $c\to \infty$ and
			$c\rightarrow 0$, respectively. In the galilean limit
			$c\to\infty$, the lightcone opens up and Einstein's
			principle of relativity gives way to the well-known galilean
			relativity principle. In the carrollian limit $c\to 0$ on
			the other hand, the lightcone closes up, implying that
			physical entities only move along the time direction. In the
			intermediate regimes where $c\ll 1$ and $c\gg 1$, physical
			systems are described by expansions in either $c$ or $1/c$
			(see,
			e.g.,~\cite{Hansen:2020pqs,Hansen:2021fxi,Hartong:2023yxo,Hartong:2023ckn}).}
		\label{fig:lightcones}
	\end{figure}
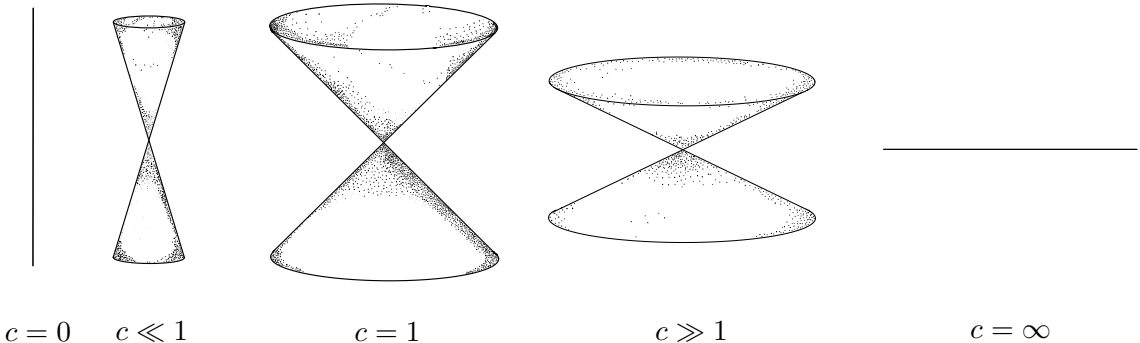

	Originally obtained as the vanishing speed-of-light limit of the
	Poincaré group~\cite{Levy1965,SenGupta1966OnAA}, carrollian
	symmetries have since played a role in the description of a wide array
	of physical phenomena, ranging from black hole
	physics~\cite{Penna:2018gfx,Donnay:2019jiz,Redondo-Yuste:2022czg,Bagchi:2023cfp}
	to exotic condensed matter systems~\cite{Bidussi:2021nmp,
		Marsot:2022imf,Bagchi:2022eui,
		Figueroa-OFarrill:2023vbj,Figueroa-OFarrill:2023qty}. Concretely,
	the Carroll algebra may be obtained by contracting the Poincaré
	algebra by reinstating factors of the speed of light $c$ in the
	Minkowski metric that appears in the commutation relations, and then
	taking the limit $c\to 0$ (see Fig.~\ref{fig:lightcones}). The
	resulting algebra has the same number of generators as the original
	Poincaré algebra. In $(d+1)$ spacetime dimensions the nonzero
	brackets of the Carroll algebra are
	\begin{equation}
	\label{eq:Carroll-algebra}
	\begin{split}         
	[\mathsf{J}_{ab},\mathsf J_{cd}] &= \delta_{bc} \mathsf J_{ad} - \delta_{ac} \mathsf J_{bd} - \delta_{bd} \mathsf J_{ac} + \delta_{ad} \mathsf J_{bc}\,,\\
	[\mathsf J_{ab}, \mathsf P_c] &= \delta_{bc} \mathsf P_a - \delta_{ac} \mathsf P_b\,,\\
	[\mathsf J_{ab}, \mathsf C_c] &= \delta_{bc} \mathsf C_a - \delta_{ac} \mathsf C_b\,,\\
	[\mathsf C_a,\mathsf P_b] &= \delta_{ab} \mathsf H\,,
	\end{split}
	\end{equation}
	where $a=1,\dots,d$ is a spatial index, and where
	$(\mathsf H,\mathsf P_a)$ generate spacetime translations,
	$\mathsf C_a$ are the generators of carrollian boosts and
	$\mathsf J_{ab}$ are the generators of $d$-dimensional spatial
	rotations. In the same way that a lorentzian geometry locally realises
	Poincar\'e symmetry, a carrollian geometry locally realises Carroll
	symmetry.\footnote{For a different notion of carrollian geometry see,
		e.g.,~\cite{Ciambelli:2018xat,Ciambelli:2018wre}. Similarities and
		differences between their approach and the perspective adopted here
		were discussed in~\cite[Sec.~2.6]{Baiguera:2022lsw}.} In a
	carrollian geometry, the notion of a metric gets replaced by a
	carrollian structure, which consists of a vector field and a
	degenerate spatial ``metric''.
	
	A fundamental ingredient of holography is that states in the Hilbert
	space of the theory can be carried both by fields living in the
	spacetime bulk and by conformal fields living at the spacetime
	conformal boundary. Indeed, massless particle states can be encoded
	into massless fields in $(d+1)$-dimensional Minkowski space
	$\mathbb{M}_{d+1}$ \textit{or} into carrollian conformal fields on
	$\scri_{d} \cong \mathbb{R} \times \mathbb{CS}^{d-1}$, a simple fact
	underlying the very foundation of carrollian
	holography~\cite{Banerjee:2018gce,Donnay:2022wvx,Nguyen:2023vfz}.
	However, such carrollian conformal fields cannot possibly carry
	massive particle states as already pointed out
	in~\cite{Nguyen:2023vfz}, since the latter propagate to/from timelike
	infinity $i^\pm$ rather than null infinity $\scri$. Hence it is
	natural to look for a holographic description of massive particles at
	timelike infinity, or more precisely on the geometry which
	appropriately captures the relevant structure near timelike infinity.
	The two points $i^\pm$ in the conformal compactification of Penrose
	obviously do not provide enough structure to properly describe massive
	particle states. We will argue that the appropriate geometric
	structure is that of
	$\zTi_{d+1} \cong \mathbb{R} \times
	\HH^d$~\cite{Figueroa-OFarrill:2021sxz}, or equivalently carrollian
	anti-de
	Sitter~\cite{Figueroa-OFarrill:2018ilb,Morand:2018tke,Figueroa-OFarrill:2019sex},
	which is a curved carrollian manifold of the same dimensionality as
	the spacetime bulk $\mathbb{M}_{d+1}$ itself. The space $\zTi_{d+1}$
	is the blowup of $i^\pm$ in perfect analogy with the Ashtekar--Hansen
	structure used for spatial infinity $i^0$
	\cite{Ashtekar:1978zz,Ashtekar1980,Gibbons:2019zfs,Figueroa-OFarrill:2021sxz}.
	The geometry of $\zTi_{d+1}$ can be viewed as a fibration over the
	hyperbolic space $\HH^d$, which will end up playing the role of the
	momentum mass shell for massive particles. While $\HH^d$ admits a
	natural action of the Lorentz group $\SO(d,1)$ as its isometry group,
	the additional null direction along the fibres allows for a nontrivial
	representation of the abelian translation subgroup $T^{d+1}$ as well.
	In this work we will develop an intrinsic description of $\zTi_{d+1}$
	and the associated carrollian fields encoding massive particle states,
	and subsequently demonstrate that these indeed arise when considering
	massive fields in Minkowski spacetime in their limit to timelike
	infinity. This will provide an ``extrapolate dictionary'', by which we
	mean that carrollian fields on $\zTi_{d+1}$ are obtained by taking the
	late time asymptotic limit of massive fields in $\mathbb{M}_{d+1}$.
	Conversely we will also provide an algorithm to reconstruct the
	massive fields in $\mathbb{M}_{d+1}$ from the corresponding carrollian
	fields on $\zTi_{d+1}$. 
	
	Let us note that the notion of holography is usually reserved for a
	description of spacetime physics through a lower-dimensional
	theory. At first sight what we describe in this work is therefore not
	strictly holographic in this sense. But even though there is no
	dimensional reduction in going from $\mathbb{M}_{d+1}$ to
	$\zTi_{d+1}$, the dependence of the massive degrees of freedom on the
	``time'' fibre $\mathbb{R}$ over $\HH^d$ is essentially trivial as
	will be made explicit in this work. More specifically, the
	time-dependence of the carrollian fields is simply fixed by the
	corresponding mass. The true degrees of freedom are obtained by
	factoring out this time-dependence, and will be shown to exactly
	correspond to the particle degrees of freedom on their momentum
	mass-shell identified with the base space $\HH^d$. From that
	perspective what we describe is holographic, in that the true degrees
	of freedom live on a codimension one surface $\HH^d$ located in some
	asymptotic region of $\mathbb{M}_{d+1}$. Another way to put it is that
	the momentum mass-shell $\mathbb{H}^d$ of the massive particles can be
	embedded into $\zTi_{d+1}$, which itself captures the physics
	happening in the asymptotic future of $\mathbb{M}_{d+1}$.  Scattering
	amplitudes are defined on the momentum mass-shell $\mathbb{H}^d$,
	which is codimension-one with respect to $\mathbb{M}_{d+1}$, and can
	therefore be viewed as the local observables of a dual theory.
	
	\begin{figure}[t!]
		\centering
		\begin{tikzpicture}[scale=0.7]
		%\foreach \Escala [count=\xi] in {1,0.8,...,0.4}
		%\node[starburst, scale=\Escala, fill=colour\xi, minimum width=1cm, minimum height=1cm, line width=1.5pt]at (6.5,0) {};
		\foreach \Escala [count=\xi] in {1,0.8,...,0.4}
		\node[starburst, scale=\Escala, fill=colour\xi, minimum width=1cm, minimum height=1cm, line width=1.5pt]at (1.2,5.5) {};
		\begin{pgfonlayer}{nodelayer}
		\node [style=dot] (0) at (0, 5) {};
		\node [style=dot] (1) at (0, -5) {};
		\node [style=dot] (2) at (5, 0) {};
		\node [style=dot] (3) at (-5, 0) {};
		\node [style=ghost] (4) at (2.5, 2.5) {};
		\node [style=ghost] (5) at (-2.5, 2.5) {};
		\node [style=ghost] (6) at (-2.5, -2.5) {};
		\node [style=ghost] (7) at (2.5, -2.5) {};
		\node [style=ghost] (8) at (0, 0) {};
		\node [style=ghost] (9) at (3, 3) {$\scri^+_d$};
		\node [style=ghost] (10) at (-3, 3) {$\scri^+_d$};
		\node [style=ghost] (11) at (3, -3) {$\scri^-_d$};
		\node [style=ghost] (12) at (-3, -3) {$\scri^-_d$};
		\node [style=ghost] (13) at (0.1, 5.5) {$i^+$};
		\node [style=ghost] (14) at (5.5, 0) {$i^0$};
		\node [style=ghost] (15) at (-5.5, 0) {$i^0$};
		\node [style=ghost] (16) at (0.1, -5.5) {$i^-$};
		\node [style=ghost] (17) at (6, 0.25) {};
		\node [style=ghost] (18) at (6, -0.25) {};
		\node [style=ghost] (19) at (7.5, 1) {};
		\node [style=ghost] (20) at (7.5, -1) {};
		\node [style=ghost] (21) at (0.75, 5.75) {};
		\node [style=ghost] (22) at (0.75, 5.25) {};
		\node [style=ghost] (23) at (2.25, 6.5) {};
		\node [style=ghost] (24) at (2.25, 4.5) {};
		\node [style=ghost] (25) at (3.3, 6.8) {$\zTi_{d+1}\cong \zAdSC_{d+1}$};
		\node [style=ghost] (26) at (3, 4.5) {{\color{red}$\mathbb{H}^d$}};
		\node [style=ghost] (27) at (3, 6.5) {};
		\node [style=ghost] (28) at (3, 5) {};
		\end{pgfonlayer}
		\begin{pgfonlayer}{edgelayer}
		\draw [style=thick black line] (0.center) to (3.center);
		\draw [style=thick black line] (0.center) to (2.center);
		\draw [style=thick black line] (3.center) to (1.center);
		\draw [style=thick black line] (1.center) to (2.center);
		\draw [style=thick black line] (5.center) to (7.center);
		\draw [style=thick black line] (4.center) to (6.center);
		\draw [style=red line, bend left, looseness=1.25] (6.center) to (7.center);
		\draw [style=red line, bend right, looseness=1.25] (6.center) to (7.center);
		\draw [style=red line, bend right=45, looseness=1.75] (6.center) to (7.center);
		\draw [style=red line, bend left=45, looseness=1.75] (6.center) to (7.center);
		\draw [style=red line] (6.center) to (7.center);
		\draw [style=red line, bend right=45] (5.center) to (4.center);
		\draw [style=red line, bend left, looseness=1.25] (5.center) to (4.center);
		\draw [style=red line, bend right=45, looseness=1.75] (5.center) to (4.center);
		\draw [style=red line, bend left=45, looseness=1.75] (5.center) to (4.center);
		\draw [style=red line] (5.center) to (4.center);
		\draw [style=blue line, bend right, looseness=1.25] (5.center) to (6.center);
		\draw [style=blue line, bend left, looseness=1.25] (5.center) to (6.center);
		\draw [style=blue line, bend right=45, looseness=1.75] (5.center) to (6.center);
		\draw [style=blue line, bend left=45, looseness=1.75] (5.center) to (6.center);
		\draw [style=blue line] (5.center) to (6.center);
		\draw [style=blue line, bend left, looseness=1.25] (4.center) to (7.center);
		\draw [style=blue line, bend right, looseness=1.25] (4.center) to (7.center);
		\draw [style=blue line, bend left=45, looseness=1.75] (4.center) to (7.center);
		\draw [style=blue line, bend right=45, looseness=1.75] (4.center) to (7.center);
		\draw [style=blue line] (4.center) to (7.center);
		\draw [style=thick black line, bend left, looseness=1.25] (22.center) to (24.center);
		\draw [style=thick black line, bend right] (21.center) to (23.center);
		\draw [style=arrow] (28.center) to (27.center);
		\end{pgfonlayer}
		\end{tikzpicture}
		\caption{The blowup of future timelike infinity $i^+$ in $(d+1)$-dimensional flat spacetime $\MM_{d+1}$ is the carrollian space $\zTi_{d+1}\cong \zAdSC_{d+1}$, which has the structure of a trivial line bundle over $d$-dimensional hyperbolic space $\mathbb{H}^d$. The Penrose diagram above includes the hyperbolic foliation of flat spacetime, where red lines correspond to copies of $\mathbb{H}^d$, while blue lines correspond to $d$-dimensional de Sitter spaces.}
		\label{fig:Minkowski}
	\end{figure}
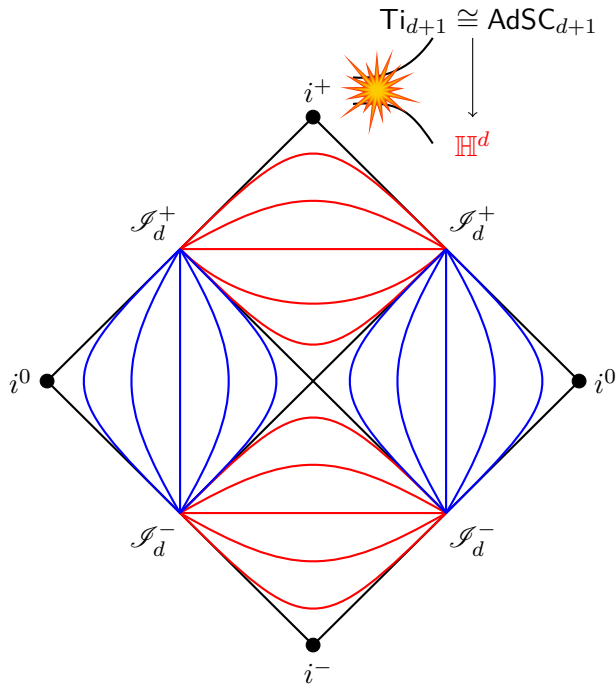
	
	Given that we are interested in giving a carrollian field theory
	description of massive particle states, i.e., unitary irreducible
	representations (UIRs) of the Poincaré group
	$\ISO(d,1)$~\cite{Wigner:1939cj}, the latter will play a central role
	in our discussion. In particular all three geometries of relevance to
	carrollian holography, namely $\mathbb{M}_{d+1}$, $\scri_d$ and
	$\zTi_{d+1}$, can be introduced as homogeneous spaces of
	$\ISO(d,1)$~\cite{Figueroa-OFarrill:2021sxz}. They can be thought of
	describing different parts of the Penrose diagram of Minkowski
	spacetime as depicted in Figure~\ref{fig:Minkowski}, and further all
	arise as Poincaré orbits in a higher-dimensional embedding space
	$\mathbb{R}^{d+1,2}$. With such homogeneous spaces at hand, we are
	confronted with the task of constructing covariant field
	representations of $\ISO(d,1)$ encoding the particle UIRs of interest.
	This can be achieved using the method of induced representations,
	inducing from a representation of the isotropy group of the
	homogeneous space, and subsequently determining the intertwining
	relation with one of the particle UIRs. The construction of carrollian
	conformal fields on $\scri_d$ carrying massless particle states of
	arbitrary spin was performed in this way in \cite{Nguyen:2023vfz}. In
	this work we will complete the story with the construction of
	carrollian fields on $\zTi_{d+1}$ encoding massive particle states of
	arbitrary spin.
	
	While bulk fields encode particle states in an intricate
	fashion, it is a generic phenomenon that boundary fields encode them
	in a much more economical way. For instance the encoding of spinning
	massless particles into fields in $\mathbb{M}_{d+1}$ requires the
	introduction of a wave equation, transversality conditions and the
	concept of gauge redundancies, while the carrollian conformal fields
	on $\scri_d$ essentially are the massless particle states up to a
	single integral transform in energy
	\cite{Banerjee:2018gce,Donnay:2022wvx,Nguyen:2023vfz}. The story we
	will present for massive particles is very similar in that respect.
	After factorising the dependence on the carrollian time direction in
	$\zTi_{d+1}$, which will be fully fixed by the mass of the
	corresponding particle, the leftover carrollian field exactly
	corresponds to the particle wavefunction on its momentum mass shell
	$\HH^d$. The extrapolate dictionary alluded to above can therefore be
	summarised in physical terms as follows: the momentum wavefunctions of
	the asymptotic particle states are directly obtained by taking the
	limit to $i^\pm$ of the massive fields in $\mathbb{M}_{d+1}$.
	
	In a nutshell the carrollian field theory on $\zTi_{d+1}$ very
	naturally describes the physics of asymptotic massive particles. This
	will be exemplified with the soft graviton theorem
	\cite{Weinberg:1965nx} and its formulation as the Ward identity
	associated with BMS supertranslations
	\cite{Strominger:2013jfa,He:2014laa,Campiglia:2015kxa}. In practice
	this formulation involves the identification of canonical charges
	generating the supertranslations of the various particles involved in
	the gravitational scattering process of interest. We will show that
	the charge which generates the supertranslation of the scattered
	massive particles \cite{Campiglia:2015kxa} is a standard Noether
	charge from the perspective of $\zTi_{d+1}$. Indeed the algebra of
	vector fields preserving the carrollian structure on $\zTi_{d+1}$ is
	actually much bigger than the Poincaré algebra, and in particular
	contains symmetries that may be identified with BMS supertranslations.
	Hence a Noether current is easily constructed by contracting a
	suitable notion of a carrollian energy-momentum tensor on $\zTi_{d+1}$
	with a supertranslation Killing vector field on $\zTi_{d+1}$, and the
	corresponding conserved charge will be shown to exactly coincide with
	the canonical generators considered in \cite{Campiglia:2015kxa} in
	relation to the subleading soft graviton theorem. Part of the analysis
	will consist in giving a full characterisation of what is meant by a
	carrollian energy-momentum tensor on $\zTi_{d+1}$.
	
	\paragraph*{Outline of the paper}
	\label{sec:outline}
	We now give a brief outline of the paper, illustrating some of the key
	results by using massive scalar particles and fields.
	
	In Section~\ref{sec:ads-carr-intr} we introduce the carrollian
	geometry of timelike infinity. In particular, after we provide a short
	introduction to carrollian geometry
	(Section~\ref{sec:carrollian-geom}), we describe how $\zTi=\zAdSC$
	arises as a homogeneous space of the Poincaré group
	(Section~\ref{sec:mink-ti}) with carrollian invariants
	(Section~\ref{sec:carroll-struc-Ti}). In Section~\ref{sec:PoinTiem} we
	also provide a useful embedding space picture.
	
	Using these symmetries we construct in
	Section~\ref{sec:massive-reps-and-fields-at-Ti} massive carrollian
	fields of generic integer spin living on $\zTi_{d+1}$, inducing from
	an irreducible representation of the isotropy group of $\zTi_{d+1}$.
	We explain that they can be understood as reducible unitary
	representations of the Poincaré group and we show that imposing
	irreducibility by fixing the quadratic Casimir directly reproduces
	Wigner's irreducible representations.
	
	In Section~\ref{sec:relat-fields-mink} we connect these intrinsically
	defined carrollian fields to the late time limit of massive fields in
	Minkowski space (Section~\ref{sec:connection-with-bulk}). Considering
	a hyperbolic foliation of Minkowski space such that
	$\eta=-d\tau+\tau^{2}h_{\HH}$ with $h_{\HH}$ the metric of a
	$d$-dimensional hyperboloid $\mathbb{H}_d$, we identify
	$(\partial_\tau,h_{\HH})$ as the appropriate carrollian structure as $\tau \to \infty$. We
	then provide an \textit{extrapolate dictionary} relating massive
	fields $\phi$ in $\mathbb{M}_{d+1}$ to carrollian fields $\bar \phi$
	in $\zTi_{d+1}$, which in the spinless case simply reads
	\begin{equation}
	\bar \phi(\tau,\x)\overset{\tau \to \infty}{\sim} \tau^{\frac{d}{2}} \phi(\tau,\x)\,, \qquad (\partial_\tau^2+m^2)\, \bar \phi(\tau,\x)=0\,.
	\end{equation}
	In particular we note that the second equation, inferred from the
	Klein--Gordon equation, amounts to an irreducibility condition as it
	fixes the quadratic Casimir operator $\mathcal{C}_2=-\partial_\tau^2$
	acting on the carrollian field representation $\bar \phi$ on
	$\zTi_{d+1}$. As a result the carrollian field can be decomposed as
	\begin{equation}
	\bar \phi(\tau,\x)=\phi^+(\x)\, e^{im\tau}+\phi^-(\x)\, e^{-im\tau} \,.
	\end{equation}
	In Section~\ref{sec:relation-plane-waves} we use a stationary phase
	approximation on the plane wave expansion of the minkowskian field
	$\phi$ to demonstrate that the carrollian modes $\phi^\pm$ are simply
	the momentum particle operators,
	\begin{equation}
	\phi^{+}(\x)=\frac{m^{d/2-1}}{2 (2\pi)^{d/2}}\, a_k\,, \qquad 
	\phi^{-}(\x)=\frac{m^{d/2-1}}{2 (2\pi)^{d/2}}\, a^{\dagger}_k\,,
	\end{equation}
	where the implicit relation $k(\x)$ identifies the asymptotic
	hyperboloid $\HH^d$ to be the momentum mass shell of the corresponding
	particles. In Section~\ref{sec:relat-mass-celest} we connect the
	carrollian fields to the celestial conformal primary
	basis~\cite{Pasterski:2016qvg}, providing us with a triangular
	relation between massive momentum, carrollian and celestial states,
	cf.~Figure~\ref{fig:triangle}.
	
	In Section~\ref{sec:EMT-and-conservation-laws} we provide an intrinsic
	definition of a carrollian energy-momentum tensor (EMT) on
	$\zTi_{d+1}$ (Section~\ref{sec:EMT-intrinsic}), which is then applied
	to the massive or ``electric'' Carroll scalar
	(Section~\ref{sec:EMT-scalar}). For the latter we also show that the
	intrinsic description agrees with the late-time limit of massive
	fields in Minkowski space and we also derive it intrinsically from the
	action of the electric scalar. In Section~\ref{sec:soft-graviton} we
	discuss the relation between the $\zTi$-supertranslations and BMS
	symmetries in connection to soft graviton theorems. In particular we
	show that the charges generating BMS transformations of massive
	particle states are a special case of Noether charges on $\zTi$,
	\begin{equation}
	Q[\xi]=\int_{\mathbb{H}^d}\varepsilon^{(d)}\,T\indices{^\tau_\mu}\xi^\mu \,,
	\end{equation}
	for an appropriate choice of Killing vector field of the form
	$\xi=S(\x) \partial_\tau$ on $\zTi$, and with $T\indices{^\mu_\nu}$
	the conserved carrollian EMT. Hence an obvious advantage of the $\zTi$
	framework is that BMS symmetries are naturally incorporated as part of
	the kinematic symmetries, just like it happens at null infinity
	$\scri$.
	
	In Section~\ref{sec:outlook} we discuss further avenues of
	investigation. In particular we argue that scattering amplitudes of
	massive particles can be identified with carrollian correlation
	functions on $\zTi$ since they satisfy the required Ward identities.
	We also introduce natural definitions of detector operators measuring,
	for instance, the total energy density carried by scattered massive
	particles along any given direction from the origin of the detector,
	and which constitute the massive counterparts of ``energy-flow''
	operators usually defined at null infinity $\scri$. We also comment on
	the inclusion of superrotations and the generalisation to curved
	space.

	\section{The carrollian geometry of timelike infinity}
	\label{sec:ads-carr-intr}
	
	In this section, we provide a brief review of carrollian geometry. We
	then recall the description of the blowup of timelike infinity as a
	homogeneous carrollian geometry. First we review its construction as a
	homogeneous space of the Poincaré group and contrast it with
	Minkowski space, followed by a description using the embedding space
	formalism.
	
	\subsection{Carrollian geometry in a nutshell}
	\label{sec:carrollian-geom}
	
	A carrollian geometry is an example of a ``non-lorentzian
	geometry'', where the familiar notion of a metric is replaced with
	something else: in this case, the metric is replaced with a
	\textit{carrollian structure} that realises local Carroll
	symmetry~\eqref{eq:Carroll-algebra}, instead of the local Poincaré
	symmetry that characterises lorentzian geometries.
	
	There are many ways to arrive at carrollian geometry. For example, one
	may obtain a carrollian geometry by formally expanding a lorentzian
	geometry in powers of the speed of
	light~\cite{Hansen:2021fxi,Armas:2023dcz}. Another approach involves
	``gauging'' the Carroll algebra~\eqref{eq:Carroll-algebra}, which from
	a mathematical perspective corresponds to the construction of a Cartan
	geometry modelled on flat carrollian spacetime considered as a
	homogeneous space of the Carroll
	group~\cite{Hartong:2015xda,Figueroa-OFarrill:2022mcy}.
	
	Regardless of its origin, a carrollian geometry is a
	$(d+1)$-dimensional manifold $\mathcal{M}$ equipped with a carrollian
	structure consisting of a nowhere-vanishing vector field
	$n = n^\mu \D_\mu$ and a symmetric tensor
	$q = q_{\mu\nu}dx^\mu dx^\nu$, where $\{x^\mu\}_{\mu=0,\dots,d}$ is a
	chart on $\mathcal{M}$. The carrollian vector field $n$ is in the
	kernel of the ``spatial metric'' $q$, namely
	\begin{equation}
	\label{eq:orthogonality}
	n^\mu q_{\mu\nu} = 0\,.
	\end{equation}
	It is useful to introduce ``inverses'' to the carrollian structure
	given by $\tau = \tau_\mu dx^\mu$ and
	$\gamma = \gamma^{\mu\nu}\D_\mu\D_\nu$, which satisfy the relations
	\begin{equation}
	\label{eq:carroll-rels}
	\gamma^{\mu\nu}\tau_\nu = 0\,,\qquad \tau_\mu n^\mu = 1\,,\qquad \delta^\mu_\nu = \gamma^{\mu\rho}q_{\rho\nu} + n^\mu\tau_\nu\,.
	\end{equation}
	In the same way that a pseudo-riemannian (or lorentzian) geometry
	locally realises Poincaré symmetry, a carrollian geometry locally
	realises Carroll symmetry. Specifically, just like the vielbeins that
	make up the metric transform under local Lorentz transformations, the
	``inverses'' $\tau$ and $\gamma$ transform infinitesimally under local
	Carroll boosts as
	\begin{equation}
	\label{eq:boost-trafos}
	\delta_C \tau_\mu = \lambda_\mu\,,\qquad \delta_C \gamma^{\mu\nu} = -2\lambda_\rho \gamma^{\rho(\mu}n^{\nu)}\,,
	\end{equation}
	where $\lambda_\mu$ is the parameter of an infinitesimal Carroll boost satisfying $\lambda_\mu n^\mu = 0$. 
	
	Another important geometric ingredient is the adapted affine
	connection $\nabla$ that defines the covariant derivative. This
	covariant derivative is compatible with the carrollian structure in
	the sense that
	\begin{equation}
	\label{eq:15}
	\nabla_\mu n^\nu=0\,,\qquad \nabla_\lambda q_{\mu\nu}=0\,.
	\end{equation}
	In general, $\nabla$ has nonzero torsion and, in contrast to
	connections in lorentzian geometries, it is not Carroll boost
	invariant unless the ``carrollian spin connection'' is included as an
	independent degree of freedom (see,
	e.g.,~\cite{Hartong:2015xda,Figueroa-OFarrill:2020gpr,Figueroa-OFarrill:2022mcy,Armas:2023dcz}).
	However, when the carrollian geometry is torsion-free, the (boost
	invariant) affine connection is determined in terms of the carrollian
	structure~\cite{Hartong:2015xda,Bekaert:2015xua}
	\begin{equation}
	\label{eq:torsion-free-connection}
	\Gamma^\rho_{\mu\nu} = n^\rho\D_{(\mu}\tau_{\nu)} + \frac{1}{2}\gamma^{\rho\lambda} (\D_\mu q_{\lambda\nu} + \D_\nu q_{\lambda\mu} - \D_\lambda q_{\mu\nu}) - n^\rho\Xi_{\mu\nu}\,,
	\end{equation}
	where $\Xi_{\mu\nu} = \Xi_{\nu\mu}$ is an arbitrary symmetric tensor
	satisfying $n^\mu\Xi_{\mu\nu}$. Parenthetically, we remark that when a
	carrollian geometry is obtained by expanding a lorentzian geometry in
	powers of the speed of light, the resulting connection constructed
	from the expansion of the Levi-Civita connection is not boost
	invariant, though it is compatible with the carrollian
	structure~\eqref{eq:15} \cite{Hansen:2021fxi}. Finally, for later use
	we introduce the natural notion of a carrollian Killing vector
	$\xi^\mu$ as that which preserves the carrollian structure, i.e.,
	\begin{equation}
	\label{eq:carrollian-KVF}
	\pounds_\xi n^\mu = \pounds_{\xi}q_{\mu\nu} = 0\,,
	\end{equation}
	where $\pounds$ denotes the Lie derivative. 
	
	\subsection{Two homogeneous spaces of the Poincaré group: $\MM$ and $\zTi$}
	\label{sec:mink-ti}
	
	It was shown in~\cite{Figueroa-OFarrill:2021sxz} that the blowup of
	timelike infinity $\zTi$ in the sense of
	Ashtekar--Hansen~\cite{Ashtekar:1978zz} is described in terms of
	carrollian anti-de Sitter space
	$\zAdSC$~\cite{Figueroa-OFarrill:2018ilb,Morand:2018tke,Figueroa-OFarrill:2019sex}.
	This space arises, among other realisations, as a homogeneous space of
	the Poincaré group, which in $(d+1)$ dimensions is
	$\ISO(d,1) \cong \SO(d,1)\ltimes \RR^{d+1}$, where $\SO(d,1)$ is the
	Lorentz group and $\RR^{d+1}$ denotes the spacetime translations. In
	contrast to the construction of $(d+1)$-dimensional Minkowski space
	$\MM_{d+1}$ as homogeneous space, where the stabiliser is the Lorentz
	group $\SO(d,1)$, the stabiliser of $\zTi_{d+1}\cong \zAdSC_{d+1}$ is
	$\ISO(d)$ consisting of the rotations and the spatial translations,
	i.e.,
	\begin{align}
	\label{eq:mink-adsC}
	\MM_{d+1} \cong \frac{\ISO(d,1)}{\SO(d,1)}\,,\qquad  \zTi_{d+1}\cong  \frac{\ISO(d,1)}{\ISO(d)} \, .
	\end{align}
	Infinitesimally, these homogeneous spaces are captured by a Klein pair
	$(\g,\h)$, where $\g $ is the Poincaré Lie algebra and $\h$ is the Lie
	algebra of the stabiliser subgroup. In a basis consisting of Lorentz
	transformations $\so(d,1)=\braket{J_{\mu\nu}}$ and spacetime translations
	$\mathfrak{t}=\braket{P_\mu}$, where
	$\mu,\nu = 0,\dots,d$, the nonzero brackets of the Poincaré
	algebra $\g$ are
	\begin{align}
	\label{eq:poincare-alg-orig}
	\begin{split}
	[J_{\mu\nu}, J_{\rho\sigma}] &= \eta_{\nu \rho} J_{\mu\sigma} -
	\eta_{\mu\rho} J_{\nu\sigma} - \eta_{\nu\sigma} J_{\mu\rho} + \eta_{\mu\sigma} J_{\nu\rho}\,, \\
	[J_{\mu\nu},P_\rho] &= \eta_{\nu\rho} P_\mu - \eta_{\mu\rho} P_\nu\,.    
	\end{split}
	\end{align}
	Note that $\mathfrak{t}$ forms an ideal of $\g$ since
	$[\g,\mathfrak{t}] \subset \mathfrak{t}$. It is useful to decompose
	the generators of the Poincaré algebra by splitting the indices
	into temporal and spatial directions $\mu=(0,a)$, where
	$a = 1,\dots,d$, and writing $B_a = J_{0a}$ and $H=P_0$. In terms of
	these generators, the Poincaré algebra~\eqref{eq:poincare-alg-orig}
	becomes
	\begin{equation}\label{eq:poincare-alg}
	\begin{aligned}
	[J_{ab},J_{cd}] &= \delta_{bc} J_{ad} - \delta_{ac} J_{bd} - \delta_{bd} J_{ac} + \delta_{ad} J_{bc}\,,\\
	[J_{ab}, B_c] &= \delta_{bc} B_a - \delta_{ac} B_b\,,\\
	[J_{ab}, P_c] &= \delta_{bc} P_a - \delta_{ac} P_b\,,
	\end{aligned}
	\qquad\qquad
	\begin{aligned}
	[B_a, P_b] &= \delta_{ab} H\,,\\
	{[H,B_a]} &{= - P_a}\,,\\
	{[B_a, B_b]} & {=J_{ab}}\,.
	\end{aligned}
	\end{equation}
	Comparing this with the Carroll algebra~\eqref{eq:Carroll-algebra}, we
	observe that upon excluding the last two brackets on the right-hand
	side, and identifying $B_a = \mathsf P_a$, $P_a = \mathsf C_a$ and
	$H = -\mathsf H$, these algebras become identical (see also~\cite[\S
	4.3]{Figueroa-OFarrill:2021sxz}). The last two brackets on the
	right-hand side in terms of the ``Carroll generators'' become
	\begin{equation}
	[\mathsf H,\mathsf P_a] = \mathsf C_a\,,\qquad [\mathsf P_a,\mathsf P_b] = \mathsf J_{ab}\,,
	\end{equation}
	which are identical in form to the additional brackets that
	distinguishes the isometry algebra of anti-de Sitter space $\so(d,2)$
	from the Poincaré algebra $\iso(d,1)$. In other words, adding these
	two brackets to the Carroll algebra is analogous to the procedure of
	adding a negative cosmological constant to the Poincaré algebra, and
	therefore the algebra in~\eqref{eq:poincare-alg} can be called the
	carrollian anti-de Sitter algebra. We remark in passing that the
	$\zAdSC$ algebra also may be obtained from $c\to 0$ contraction of
	$\so(d,2)$; for additional details, we refer
	to~\cite{Figueroa-OFarrill:2018ilb}. Thus, the Klein pair of
	$\MM_{d+1}$ is $(\g,\h_{\MM})$, where
	$\mathfrak{h}_{\MM}=\langle J_{ab},B_{a}\rangle$, while the Klein pair
	of $\zTi_{d+1}$ is $(\g,\h_{\zTi})$, where
	$\h_{\zTi}=\langle J_{ab},P_{a}\rangle$. Since the stabiliser
	subalgebra $\mathfrak{h}$ in either case does not contain any
	nontrivial ideal, the homogeneous spaces are both effective. This
	means that all Poincaré transformations act non-trivially on the
	respective spacetimes.\footnote{This can be contrasted with the lower
		dimensional hyperbolic spaces $\HH^{d}$ where
		$\mathfrak{h}_{\HH}=\langle J_{ab},P_{a},H \rangle$. Since the
		algebra $\mathfrak{t} = \langle H, P_{a}\rangle$ is an ideal of
		$\mathfrak{g}$ and is contained in $\mathfrak{h}$, the time and
		spatial translations act trivially on the underlying space
		(see~\cite{Figueroa-OFarrill:2021sxz} for more
		details).\label{fn:1}} Furthermore, since both Klein pairs
	$(\mathfrak{g},\mathfrak{h})$ can be written as
	$\mathfrak{g}=\mathfrak{h}\oplus\mathfrak{m}$ with
	$[\h,\m]\subset \m$, both spaces are reductive.
	
	At this stage, one may wonder what makes the space $\zTi_{d+1}$
	carrollian. There is a simple way to derive the invariant structure of
	a homogeneous space from its algebraic description in terms of the
	Klein pair (see for example~\cite{Figueroa-OFarrill:2018ilb}): if
	$M \cong G/H$ is a homogeneous space of $G$ with reductive Klein pair
	$(\g,\h)$, the $\h$-invariant tensors on $\m$ are in one-to-one
	correspondence with $G$-invariant tensors on $M$. Concretely for
	$\zTi_{d+1}$, this means that we should look for $\h$-invariant
	tensors on the vector spaces $\m$, $\m^*$, where
	$\m^*=\text{span}(\eta,\beta^a)$ is the dual of $\m$, with
	$\eta(H) = 1$ and $\beta^a(B_b) = \delta^a_b$. The stabiliser $\h$
	acts on $\m$ via the adjoint, i.e., the bracket, while it acts on the
	dual $\m^*$ via the coadjoint action, which is defined in the
	following manner: if $\alpha \in \m^*$ and $X\in \h$, then
	$X\cdot \alpha = - \alpha \circ \ad_X$. Using this, we find that the
	brackets~\eqref{eq:poincare-alg} lead to two $\mathfrak{h}$-invariant
	tensors of low rank on $\m$, namely $H\in \m$, which corresponds to
	the $G$-invariant vector $n$ on $\zTi_{d+1}$, and
	$\delta_{ab}\beta^a\beta^b$ in the symmetric product of $\m^*$, which
	corresponds to the $G$-invariant covariant tensor $q$ on $\zTi_{d+1}$:
	precisely a carrollian structure as defined in
	Section~\ref{sec:carrollian-geom}. We will discuss the carrollian
	structure on $\zTi_{d+1}$ in more detail in
	Section~\eqref{sec:carroll-struc-Ti}.
	
	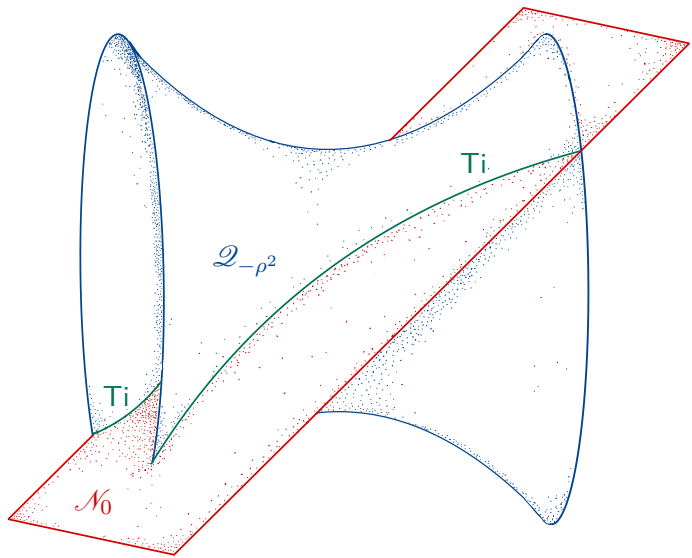
\begin{figure}[h!]
		\centering
		\definecolor{DarkRed}{RGB}{210,0,0}
		\definecolor{DarkBlue}{RGB}{3,70,143}
		\definecolor{DarkGreen}{RGB}{0,115,85}
		\includegraphics[width=0.6\textwidth]{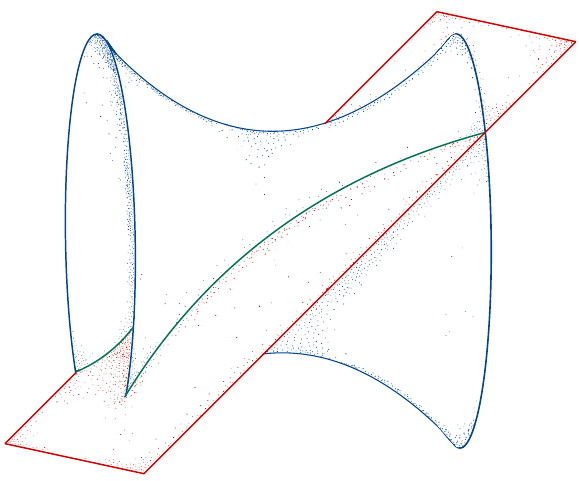}
		\begin{tikzpicture}[overlay]
		\begin{pgfonlayer}{nodelayer}
		\node [style=ghost] (0) at (-8, 0.8) {{\color{DarkRed}$\mathscr{N}_0$}};
		\node [style=ghost] (1) at (-6, 4) {{\color{DarkBlue}$\mathscr{Q}_{-\rho^2}$}};
		\node [style=ghost] (2) at (-7.7, 2.2) {{\color{DarkGreen}$\zTi$}};
		\node [style=ghost] (3) at (-3, 5.3) {{\color{DarkGreen}$\zTi$}};
		\end{pgfonlayer}
		\end{tikzpicture}
		\caption{In the embedding space $\mathbb{R}^{d+1,2}$, two copies of $\zTi$ arise as the intersection of $\mathscr{Q}_{-\rho^2}\cong \zAdS_{d+2}$ and the null hypersurface $\mathscr{N}_0$ defined by $Y^- = 0$.}
		\label{fig:Ti-embedding}
	\end{figure}
	
	\subsection{Embedding $\zTi$ into pseudo-euclidean spacetime}
	\label{sec:PoinTiem}
	
	The blowup of timelike infinity $\zTi_{d+1}$ can be embedded into
	pseudo-euclidean space $\mathbb R^{d+1,2}$: the \textit{embedding
		space}~\cite{Figueroa-OFarrill:2021sxz} (see also Appendix B
	of~\cite{Salzer:2023jqv} and~\cite{Bekaert:2022oeh}). Indeed,
	Minkowski space itself, along with a litany of other spaces such as
	the blowup of spatial infinity $\zSpi$, null infinity $\scri$, and
	even the Schwarzschild solution, can be realised as embeddings in
	$\mathbb
	R^{d+1,2}$~\cite{kasner1921finite,Fronsdal:1959zza,Penrose:1986ca,Figueroa-OFarrill:2021sxz}.
	It also plays an important role in the AdS/CFT correspondence (see for
	example~\cite{Costa:2011mg, Costa:2014kfa}) since, as we shall see
	below, $\zAdS_{d+2}$ may also be embedded into $\mathbb R^{d+1,2}$.
	
	We chart the embedding space $\mathbb{R}^{d+1,2}$ using
	\begin{equation}
	\label{eq:embedding}
	Y^A=\begin{pmatrix}y^\mu\\Y^+\\Y^-\end{pmatrix}\,,
	\end{equation}
	where $A=(\mu,+,-)$, and where $Y^\pm$ are null coordinates. In these coordinates, the metric on $\RR^{d+1,2}$ reads
	\begin{equation}
	\label{eq:metricem}
	ds^2 = \eta_{AB}\, d Y^B d Y^A=\eta_{\mu\nu}\, d y^\mu d y^{\nu}+2 d Y^+d Y^-\,,
	\end{equation}
	where $\eta_{\mu\nu}$ is the metric of $(d+1)$-dimensional Minkowski
	space with signature $(-,+,+, \cdots)$.
	Following~\cite{Figueroa-OFarrill:2021sxz}, we define the
	$(d+2)$-dimensional space $\mathscr{Q}_\Lambda$ with $\Lambda \in \RR$
	as the quadric given by $Y^A\eta_{AB}Y^B=\Lambda$. The group
	$\operatorname{O}(d+1,2)\subset \ISO(d+1,2)$ preserves these quadrics
	and acts on the embedding space via the vector fields
	\begin{equation}
	\xi_{J_{AB}} = \eta_{AC}Y^C  \D_B - \eta_{AC}Y^C\D_A\,,%\in \mathscr{X}(\mathbb{R}^{d+1,2})\,,
	\end{equation}
	which split under~\eqref{eq:embedding} as
	\begin{subequations}
		\begin{align}
		\xi_{J_{\mu\nu}} &= y_\mu \D_\nu - y_\nu \D_\mu\,,& \xi_{J_{\mu+}} &= y_\mu\D_+ - Y^-\D_\mu\,,\label{eq:first}\\
		\xi_{J_{\mu-}} &= y_\mu \D_- - Y^+ \D_\mu\,,& \xi_{J_{+-}} &= Y^- \D_- - Y^+ \D_+\,,
		\end{align}
	\end{subequations}
	where we notice that the vector fields in~\eqref{eq:first} generate
	the $(d+1)$-dimensional Poincaré algebra $\iso(d,1)$. The quadric
	$\mathscr{Q}_{\Lambda < 0}$ gives a copy of $\zAdS_{d+2}$, while
	$\mathscr{Q}_{\Lambda > 0}$ describes ``pseudo-de Sitter space'',
	which is de Sitter space with signature $(d,2)$. The null quadric
	$\mathscr{Q}_0$ can be thought of as the (generalised) lightcone of
	the embedding space. Next, we define $\mathscr{N}_\sigma$ for
	$\sigma\in \RR$ as the null surface $Y^- = \sigma$, which is preserved
	by the Poincaré group corresponding to the algebra spanned by the
	first line above for generic values of $\sigma$. We also define the
	intersection
	\begin{equation}
	\mathscr{M}_{\Lambda,\sigma} := \mathscr{Q}_{\Lambda }\cap \mathscr{N}_{\sigma}\,,
	\end{equation}
	which is preserved by the Poincaré group that preserves
	$\mathscr{N}_\sigma$. As depicted in Figure~\ref{fig:Ti-embedding},
	the blowup of timelike infinity $\zTi_{d+1}$ is then captured by
	$\mathscr{M}_{-\rho^2,0}$, where we set $\Lambda = -\rho^2$ for some
	$\rho\in \RR$ for definiteness. More precisely,
	$\mathscr{M}_{-\rho^2,0}$ breaks into two orbits under the action of
	the Poincaré group discussed above
	\begin{equation}
	\mathscr{M}_{-\rho^2,0} = \zTi_{d+1}^+\sqcup \zTi_{d+1}^-\,,
	\end{equation}
	where
	\begin{equation}
	\label{eq:2.10}
	\zTi^{\pm}_{d+1} = \left\{ \begin{pmatrix}
	y^\mu\\
	Y^+\\
	0
	\end{pmatrix}~\middle\vert ~ \eta_{\mu\nu} y^\mu y^\nu = -\rho^2\,,~~ \pm y^0 > 0\,,~~ Y^+ \in \RR  \right\}\,.
	\end{equation}
	This explicitly exhibits $\zTi_{d+1}$ as a trivial line bundle
	$\RR\times \HH^d$ over the $d$-dimensional hyperboloid $\HH^d$ defined
	by $\eta_{\mu\nu} y^\mu y^\nu = -\rho^2$, while the coordinate along
	the fibre is $Y^+$ which is to be thought of as carrollian time. For
	more details about this construction, we refer
	to~\cite{Figueroa-OFarrill:2021sxz}.
	
	Setting $\rho=1$, we now introduce coordinates $ x^{a} = \x$ via
	\begin{equation}
	\label{y to x}
	y^0 = \cosh r\,, \qquad y^a = \y= \sinh r\,\frac{x^a}{r}\,,\qquad x^a x_a=r^2\,, \qquad Y^+\equiv \tau\,,
	\end{equation}
	where the $x^a$ parameterise the hyperboloid $\HH^d$, while $\tau$ is
	now the carrollian time coordinate. Let us also introduce
	$\hat x^a = x^a/r$ which is the unit vector pointing in the direction
	of $x^a$. We will use both sets of
	coordinates $(\tau,\y)$ and $(\tau, \x)$ in the remainder of this
	paper. 
	
	In terms of the $\x$ coordinates, the Poincaré vector fields
	that preserve the embedding \eqref{eq:2.10} take the form
	\begin{subequations}
		\label{eq:fund-vector-fields}
		\begin{align}
		\xi_{J_{ab}} &=  x^a \frac{\d}{\d x^b} - x^b \frac{\d}{\d x^a}\,, \label{eq:vec-a}\\
		\xi_{B_a} &= -r \coth r \left( \frac{\d}{\d x^a} - \hat x^a \frac{\d}{\d r} \right) - \hat x^a \frac{\d}{\d r}\,, \label{eq:vec-b}\\
		\xi_H &= -\cosh r\, \frac{\d}{\d \tau}\,,\label{eq:vec-c}\\
		\xi_{P_a} &= \sinh r\, \hat x^a \frac{\d}{\d \tau} \label{eq:translation-vec}\,.
		\end{align}
	\end{subequations}
	where we used that $x^a \frac{\d}{\d x^a} = r \frac{\d}{\d r}$. As we
	shall see in what follows, these Poincaré vector fields are a subset
	of the vector fields that preserve the carrollian structure on
	$\zTi_{d+1}$.
	
	\subsection{The carrollian structure on $\zTi$}
	\label{sec:carroll-struc-Ti}
	
	Having constructed the carrollian space $\zTi_{d+1}$ both as a
	homogeneous space of the Poincaré group and by embedding it in
	$\RR^{d+1,2}$, we now turn our attention to the carrollian structure
	on $\zTi_{d+1}$. Using the methods
	of~\cite{Figueroa-OFarrill:2019sex}, one may extract the carrollian
	structure from the Maurer--Cartan form associated with $\zTi_{d+1}$.
	It is, however, simpler to derive the form of the carrollian structure
	$(n,q)$ directly from the coordinates introduced in~\eqref{y to x}:
	from~\eqref{eq:2.10}, we see how $\zTi_{d+1}\cong \RR \times \HH^d$
	has the structure of a trivial fibre bundle, where the coordinate
	along the fibre is the carrollian time, which implies that
	$n = \D_\tau$. Similarly, the carrollian ``ruler'' $q$ restricted to
	an equal-time hypersurface, i.e., a copy of $\HH^d$, is the euclidean
	metric on $\HH^d$. Hence, the carrollian structure $(n,q)$ on
	$\zTi_{d+1}$ is 
	\begin{align}
	\label{eq:invariants}
	n&=\pd_{\tau}\,, &  q &= 0  \dd\tau^{2} + h_{ab}\dd x^a dx^b \,, & h_{ab}\dd x^a \dd x^b &= \dd r^2 + \sinh^2 r~ g_{\mathbb{S}^{d-1}}.
	\end{align}
	where $h_{ab}$ is the metric on $\HH^d$ and $g_{\mathbb{S}^{d-1}}$ is
	the round metric on the unit sphere $\mathbb{S}^{d-1}$. The carrollian
	vector $n$ field is in the kernel of $q$, i.e., it satisfies the
	relation~\eqref{eq:orthogonality}. Alternatively, in the
	$\y$-coordinates introduced above the hyperbolic metric reads
	\begin{equation}
	\label{eq:invariants-2}
	h_{ab}(\y)\dd y^a \dd y^b = \left(\delta_{ab}-\frac{y_a y_b}{1+y^c y_c}\right)dy^a dy^b.
	\end{equation}
	To be able to integrate, we need the invariant volume form
	$\varepsilon$,
	\begin{subequations}
		\label{eq:volumeform}
		\begin{align}
		\varepsilon = \dd \tau\wedge \varepsilon^{(d)}&=\left(\frac{\sinh r}{r}\right)^{d-1}\frac{1}{d!}\, \varepsilon_{a_1\ldots a_d}\dd \tau \wedge\dd x^{a_1}\wedge \ldots \wedge \dd x^{a_d}\\
		&=\frac{1}{\sqrt{1+y^a y_a}}\frac{1}{d!}\, \varepsilon_{a_1\ldots a_d}\dd \tau \wedge\dd y^{a_1}\wedge \ldots \wedge \dd y^{a_d}
		\end{align}
	\end{subequations}
	where $\varepsilon^{(d)}$ is the volume form on the hyperboloid
	$\HH^d$. Furthermore, as shown in~\cite{Figueroa-OFarrill:2018ilb},
	$\zTi_{d+1}\cong \zAdSC_{d+1}$ is torsion-free, and hence we find that
	the torsion-free affine connection~\eqref{eq:15} on $\zTi_{d+1}$ has
	components
	\begin{equation}
	\label{eq:1}
	\Gamma^a_{\mu \tau}=\Gamma^{\tau}_{\tau\tau}=\Gamma^{\tau}_{a\tau}=0\,,
	\end{equation}
	while $\Gamma^a_{bc}$ is the Levi-Civita connection on the hyperboloid
	$\HH^d$ and $\Gamma^{\tau}_{ab}$ is left undetermined, corresponding
	to the arbitrary spatial tensor $\Xi_{\mu\nu}$ that appears in the
	general expression~\eqref{eq:15}.

	We may now ask what are the symmetries of the carrollian structure on
	$\zTi_{d+1}$; in other words, what is the set of vector fields
	$\xi^\mu$ that satisfy the carrollian analogue of the Killing
	equation~\eqref{eq:carrollian-KVF}? This was worked out
	in~\cite{Figueroa-OFarrill:2019sex}, and we recall the procedure
	below. Writing $\xi = \xi^\mu \D_\mu = \xi^\tau \D_\tau + \xi^a\D_a$,
	the vector $\xi$ preserves the carrollian vector field $n = \D_\tau$
	if
	\begin{equation}
	[\xi,n] = - \D_\tau \xi^\mu \D_\mu = 0\,,
	\end{equation}
	implying that $\xi^\mu = \xi^\mu(\x)$. The preservation of $q$ requires that $\pounds_\xi q = 0$, i.e.,
	\begin{equation}
	\xi^\rho \D_\rho q_{\mu\nu} + q_{\mu\rho}\D_\nu \xi^\rho + q_{\nu\rho}\D_\mu \xi^\rho = 0\,,
	\end{equation}
	but since $\xi^\mu$ does not depend on $\tau$ and $q$ has no nonzero
	$\tau$-components, the condition above reduces to
	$\pounds_{\vec \xi}\, h_{ab} = 0$, i.e., $\vec \xi := \xi^a\D_a $ must
	be a Killing vector of the metric $h_{ab}$ on $\HH^d$. The vector
	fields $\vec \xi$, as is well known, generate the Lorentz algebra
	$\so(d,1)$. Therefore, the symmetries of the carrollian structure on
	$\zTi_{d+1}$ consists of the vector fields $\xi_X$ for $X\in \so(d,1)$
	in~\eqref{eq:vec-a}--\eqref{eq:vec-b}, as well as an infinite set of
	vector fields of the form
	\begin{equation}
	\label{eq:H-bms}
	\xi_S = S(\x)\D_\tau\,,
	\end{equation}
	where $S(\x)$ is an arbitrary smooth function on $\HH^d$. In
	particular, the vector fields $\xi_{P_a}$ and $\xi_H$
	in~\eqref{eq:vec-c}--\eqref{eq:translation-vec} associated with
	Poincaré translations are special cases of~\eqref{eq:H-bms}. Hence,
	the algebra $\mathfrak{kv}(\zTi_{d+1})$ of the vector fields that are
	symmetries of the carrollian structure on $\zTi_{d+1}$ has the form
	\begin{equation}
	\label{eq:symmetries-of-Ti}
	\mathfrak{kv}(\zTi_{d+1}) = \so(d,1)\ltimes C^\infty(\HH^d) \supset \iso(d,1) \,,
	\end{equation}
	where the action of $\so(d,1)$ on $C^\infty(\HH^d)$ is the standard
	action of vector fields on functions, $[X,S] = \xi_X S$, where
	$X\in \so(d,1)$ and $f\in C^\infty(\HH^d)$. The embedding of the
	Poincaré algebra $\iso(d,1)$ into the symmetry algebra of $\zTi_{d+1}$
	is, as explained above, achieved by choosing the function $S$ to be of
	the form~\eqref{eq:vec-c}--\eqref{eq:translation-vec}.
	
	These symmetries bear a striking resemblance to the BMS symmetries,
	which arise as the asymptotic symmetries of flat
	spacetime~\cite{Bondi:1962px,Sachs:1962zza}, and which play an
	important role in celestial holography (see,
	e.g.,~\cite{Donnay:2023mrd} for a review emphasising this
	perspective). The algebra of BMS symmetries of $(d+1)$-dimensional
	asymptotically flat spacetime has the form
	\begin{equation}
	\mathfrak{bms}_{d+1} = \so(d,1)\ltimes C^\infty(\mathbb{S}^{d-1})\,,
	\end{equation}
	which differs from the symmetries of the carrollian structure on
	$\zTi_{d+1}$ in~\eqref{eq:symmetries-of-Ti} only by involving
	functions on the $(d-1)$-dimensional (celestial) sphere
	$\mathbb{S}^{d-1}$ rather than functions on $\HH^d$. In asymptotically
	flat space, a symmetry involving an arbitrary function
	$f \in C^\infty(\mathbb{S}^{d-1})$ is called a supertranslation, and,
	by analogy, we will call a symmetry of $\zTi_{d+1}$ involving an
	arbitrary function $S\in C^\infty(\HH^d)$ a
	``$\zTi$-supertranslation''. In particular, since the boundary of
	hyperbolic space $\HH^d$ is the conformal (or celestial) sphere
	$\mathbb{CS}^{d-1}$, it follows that
	\begin{equation}
	\mathfrak{kv}(\zTi_{d+1}) \supset \mathfrak{bms}_{d+1}\,,
	\end{equation}
	i.e., the BMS symmetries are a subset of the symmetries of the
	carrollian structure on $\zTi_{d+1}$. We will have more to say about
	this in Section~\ref{sec:soft-graviton}, where we discuss the
	relation to Weinberg's soft graviton theorem.
	
	The carrollian symmetries of $\zTi$ can be further restricted, e.g.,
	by adding the connection as an additional structure that needs to be
	preserved. On the other hand they can also be enlarged, similar to
	generalised BMS~\cite{Campiglia:2014yka}, by only imposing invariance
	of the vector field and the volume form~\eqref{eq:volumeform} or a
	conformal generalisation thereof (see, e.g.,~\cite{Bekaert:2022ipg}
	for a discussion of several such modifications).
	
	\section{Massive representations and carrollian fields at timelike infinity}
	\label{sec:massive-reps-and-fields-at-Ti}
	
	In this section we induce two unitary representations of the Poincaré
	group for generic spin and in generic dimension. The first are
	interpreted as carrollian fields on $\zTi$ which have to satisfy,
	analogous to the Klein--Gordon field in Minkowski space, a carrollian
	wave equation
	$(\pd_{\tau}^{2}+ m^{2})\phi_{a_1\ldots a_s}(x)=0$. We then show that this reducible
	representations can be seen as a direct integral representations of
	Wigner's unitary irreducible representation of massive
	particles~\cite{Wigner:1939cj}.
	
	\subsection{Carrollian fields on $\zTi$}
	\label{sec:carrollian-fields}
	
	We proceed with the construction of fields living on $\zTi_{d+1}$.
	This representation will be induced from the isotropy subgroup
	generated by $\mathfrak{iso}(d)=\langle P_a,J_{ab} \rangle$ and we
	will do this infinitesimally similarly to
	\cite{Mack:1969rr,Nguyen:2023vfz}. This means that the starting point
	is a finite-component unitary representation $\phi_{a_1\ldots a_s}(0)$
	of $\mathfrak{iso}(d)$,
	\begin{equation}
	\label{eq:inducing}
	\left[J_{ab},\phi_{a_1\ldots a_s}(0)\right]=\left(\Sigma_{ab}\right)^{b_1\ldots b_s}_{a_1\ldots a_s}\, \phi_{b_1\ldots b_s}(0) \,, \qquad \left[P_a,\phi_{a_1\ldots a_s}(0)\right]=0 \, ,
	\end{equation}
	where $\left(\Sigma_{ab}\right)^{b_1\ldots b_s}_{a_1\ldots a_s}$ acts
	in the corresponding tensor representation of $\mathfrak{so}(d)$.
	Although much of the discussion in this subsection will apply to any
	tensor representation of $\mathfrak{so}(d)$, including those with
	mixed-symmetry, and even to arbitrary representations of
	$\mathfrak{spin}(d)$, we will focus on real totally symmetric and
	traceless tensor representations of integer spin. In the case of the
	vector fields $\phi_{a_1}(0)$ with spin $1$ for instance, we have
	$\left(\Sigma_{ab}\right)^{b_1}_{a_1}=\delta_{b
		a_1}\delta^{b_1}_a-\delta_{a a_1}\delta^{b_1}_b$. In the following,
	we will suppress tensor indices on $\phi$ and $\Sigma$.
	
	We choose coordinates using $U : \zTi \to G$ and
	\begin{equation}
	\label{eq:coset-rep-2}
	U(\tau, \x)=e^{x^a B_a}e^{\tau H} \,,
	\end{equation}
	from which we obtain a representation of the full Poincaré group by
	translating the field to generic points on $\zTi_{d+1}$
	\begin{equation}
	\label{eq:Heisen}
	\phi(x)\equiv \phi(\tau,\x) = U(\tau, \x)\, \phi(0)\, U(\tau, \x)^{-1}\,.
	\end{equation}
	We begin by collecting the following useful relations
	\begin{align}
	\label{eq:2}
	\begin{split}
	UHU^{-1}&=e^{x^aB_a}H e^{-x^aB_a}= \exp(-\ad_{-\x\cdot \bB})H = H \cosh r+\hat x^a P_a\sinh r\,, \\
	e^{x^c B_c}B_a e^{-x^cB_c}& = -J_{ab} \hat x^b \sinh r  + \hat{x}_a\hat{x}^bB_b(1-\cosh r)+B_a\cosh r\,,\\
	UP_a U^{-1}&%e^{-x^iB_i}e^{-t H}P_ie^{t H}e^{x^iB_i}=e^{-x^iB_i}P_ie^{x^iB_i}
	=H \hat{x}_a \sinh r +\hat{x}_a\hat{x}^bP_b(\cosh r-1)+P_a\,,\\
	UJ_{ab}U^{-1}&=J_{ab}+(1-\cosh r)\hat{x}^c(\hat{x}_a J_{bc}-\hat{x}_bJ_{ac})-(\hat{x}_bB_a-\hat{x}_aB_b)\sinh r\,,\\
	\partial_a UU^{-1}&= J_{ab}\frac{\hat{x}^b}{r}(1-\cosh r)+\hat{x}_a\hat{x}^bB_b\left(1-\frac{\sinh r}{r}\right)+B_a \frac{\sinh r}{r}\,,
	\end{split}
	\end{align}
	where $r^2 = \delta_{ab}x^ax^b$ and $\hat x^a = x^a/r$ is a unit
	vector pointing in the direction of $x^a$. To derive the last equation
	we used the derivative of the exponential map
	\begin{equation}
	\label{eq:12}
	\frac{d}{d t}e^{X(t)}=e^{X(t)}\frac{1-\exp(-\ad_{X(t)} )}{\textrm{ad}_{X(t)}}\frac{d X(t)}{d t}= e^{X(t)}\sum^\infty_{k=0}\frac{(-1)^k}{(k+1)!}(\textrm{ad}_{X(t)})^k\frac{d X(t)}{d t}\,,
	\end{equation}
	where $X(t)\in C^1(\g)$ is a differentiable curve in the Lie algebra $\g$.
	
	By definition \eqref{eq:Heisen} we have
	\begin{equation}
	\left[H,\phi(0) \right]=\partial_\tau \phi(0)\,, \qquad  \left[B_a,\phi(0) \right]=\partial_a \phi(0)\,.
	\end{equation}
	From~\eqref{eq:inducing}, it follows
	immediately that
	\begin{equation}
	\label{eq:9}
	0=[P_a,\phi(0)]=[P_a, U^{-1}\phi(x) U]=U^{-1}[U P_a U^{-1},\phi(x)]U\,,
	\end{equation}
	and therefore the third line of~\eqref{eq:2} implies that
	\begin{equation}
	\label{eq:3}
	\hat{x}_a\sinh r [H,\phi(x)]+(\cosh r-1)\hat{x}_a\hat{x}^b[P_b,\phi(x)]+[P_a,\phi(x)]=0\,.
	\end{equation}
	By the same logic, we have $[UJ_{ab}U^{-1},\phi(x)] = \Sigma_{ab}\,\phi(x)$, which leads to
	\begin{align}
	\label{eq:5}
	&[J_{ab},\phi(x)]+(1-\cosh r)\hat{x}^c(\hat{x}_a [J_{bc},\phi(x)]-\hat{x}_b[J_{ac},\phi(x)]) \nonumber\\
	&\qquad\qquad\qquad\qquad\qquad\qquad-(\hat{x}_b[B_a,\phi(x)]-\hat{x}_a[B_b,\phi(x)])\sinh r=\Sigma_{ab}\, \phi(x)\,.
	\end{align}
	On the other hand, definition~\eqref{eq:Heisen} allows us to write
	\begin{align}
	\label{eq:6}
	\partial_\tau\phi(x)&=[\partial_\tau U U^{-1},\phi(x)]
	=[UHU^{-1},\phi(x)] =[H,\phi(x)]\cosh r +\hat x^a [P_a,\phi(x)]\sinh r \,.
	\end{align}
	and similarly for the spatial derivatives
	\begin{align}
	\label{eq:10}
	\partial_a\phi(x)&=[\partial_a U U^{-1},\phi(x)]%=-[e^{-x^iB_i}\partial_ie^{x^iB_i},\phi(x)]\nonumber
	\\
	&=\frac{\hat{x}^b}{r}(1-\cosh r)[J_{ba},\phi(x)]+\hat{x}_a\hat{x}^b\left(1-\frac{\sinh r}{r}\right)[B_b,\phi(x)]+ \frac{\sinh r}{r}[B_a,\phi(x)]\,.\nonumber
	\end{align}
	We can now solve this system for the unknown brackets between the
	generators and $\phi(x)$. Contracting~\eqref{eq:3} with $\hat{x}_a$
	and combining the result with~\eqref{eq:6} yields
	\begin{equation}
	\label{eq:4}
	[H,\phi(x)]=\cosh r\, \partial_\tau\phi(x)\,,\qquad  [P_a,\phi(x)]=-\sinh r\,\hat{x}_a\partial_\tau\phi(x)\,.
	\end{equation}
	Next contracting~\eqref{eq:10} and~\eqref{eq:5} with $\hat x^a$ and
	plugging into \eqref{eq:10} and \eqref{eq:5}, we find
	\begin{equation}
	\label{eq:13}
	[B_a,\phi(x)]=\frac{r \cosh r}{\sinh r}\left(\partial_a \phi(x)-\hat{x}_a\hat{x}^b\partial_b\phi(x)\right)+\hat{x}_a\hat{x}^b\partial_b\phi(x)-\tanh(r/2) \hat{x}^b\Sigma_{ba}\, \phi(x)\,.
	\end{equation}
	and
	\begin{equation}
	\label{eq:14}
	[J_{ab},\phi(x)]=x_b\partial_a\phi(x)-x_a\partial_b\phi(x)+\Sigma_{ab}\, \phi(x)\,.
	\end{equation}
	
	In summary, the infinitesimal action of the Poincaré group on the
	$\zTi$ field $\phi(x)$ is
	\begin{subequations}
		\label{eq:action-on-fields}
		\begin{align}
		[J_{ab},\phi(x)]&=-\xi_{J_{ab}}\phi(x)+\Sigma_{ab}\, \phi(x)\,, \\
		[B_a,\phi(x)]&=-\xi_{B_{a}}\phi(x)-\tanh(r/2) \hat{x}^b\Sigma_{ba}\, \phi(x)\,,\\ 
		[H,\phi(x)] &= -\xi_{H}\phi(x)\,,\\
		[P_a,\phi(x)]&=-\xi_{P_{a}}\phi(x)\,,
		\end{align}
	\end{subequations}
	where the vector fields $\xi_X$ for $X\in \iso(d,1)$ are explicitly
	given in~\eqref{eq:fund-vector-fields}.
	
	The action of the Casimir $\mathcal{C}_2=P^{2}=-H^2+P^a P_a$ on the
	$\zTi$ fields is then given by
	\begin{align}
	\label{eq:casimir-fields}
	\left[\mathcal{C}_2,\phi(x)\right] = -\pd_{\tau}^{2}\phi(x)\,.
	\end{align}
	The Casimir equation $\mathcal{C}_2=m^2$ defining massive
	representations\footnote{The positivity of the Casimir
		$\mathcal{C}_2=m^2>0$ comes from our choice of working with
		anti-hermitian translation operators $(H,P_a)$.} then reduces to the
	simple differential equation
	\begin{align}
	\label{eq:reducibility}
	\pd_{\tau}^{2}\phi(x)   =- m^{2}\phi(x) \, .
	\end{align}
	This equation of motion, which is true for any spin, ensures that our
	fields on $\zTi_{d+1}$ describe massive Poincaré particles. It is the
	$\zTi$ analogue of the relativistic equation of motion satisfied by
	massive fields in Minkowski space such as the Klein--Gordon equation
	(see~\eqref{wave equation} below). This construction is analogous to
	\emph{massive carrollian fields}~\cite{Figueroa-OFarrill:2023qty},
	which can carry any spin, with the main difference that they are
	valued on a flat carrollian manifold, which can be seen as the flat
	limit of $\zTi$.
	
	Equation~\eqref{eq:reducibility} can be integrated to an action
	$S[\phi]$ on $\zTi_{d+1}$
	\begin{equation}
	\label{eq:free-higher-spin-action}
	S[\phi] = \frac{1}{2}\int_{\zTi_{d+1}}\varepsilon\,\left[\D_\tau \phi^{a_1\dots a_n}\D_\tau \phi_{a_1\dots a_n} - m^2 \phi^{a_1\dots a_n}\phi_{a_1\dots a_n} \right]\,,
	\end{equation}
	where $\varepsilon$ is the volume form \eqref{eq:volumeform} on
	$\zTi_{d+1}$. Due to the absence of gradient terms of the form
	$\pd_{i}\phi\pd^{i}\phi$ this theory is an example of an ultralocal
	field theory that have been studied in the 60s and 70s (see,
	e.g.,~\cite{Klauder:2000ud} and references therein) and have recently
	resurfaced in carrollian physics. When we restrict to a scalar this
	theory is the $\zTi_{d+1}$ analogue of the massive \textit{electric
		Carroll scalar theory}~\cite{Henneaux:2021yzg,deBoer:2021jej}. We
	will have more to say about this in Section~\ref{sec:EMT-scalar},
	where we explicitly consider aspects of the electric Carroll scalar
	theory on $\zTi_{d+1}$.
	
	Reinstating tensor indices, the
	solutions of~\eqref{eq:reducibility} are given by
	\begin{align}
	\label{eq:solution}
	\phi_{a_1\ldots a_s}(\tau,\x) &= \phi^{+}_{a_1\ldots a_s}(\x)\, e^{im\tau} + \phi^{-}_{a_1\ldots a_s}(\x)\, e^{-im\tau}\,,
	\end{align}
	where since $\phi$ is real, we have that $\phi^+ = (\phi^-)^*$. In
	Section~\ref{sec:connection-with-bulk} we will connect these
	carrollian fields to the expansion of massive fields in Minkowski
	space near timelike infinity. 
	But first we will elucidate on the connection to Wigner's massive
	unitary irreducible representations.
	
	\subsection{Relation to massive UIRs of the Poincaré group}
	\label{sec:massive-reps}
	
	It is useful to recall Wigner's construction of the massive unitary
	irreducible representation (UIR) of the Poincaré
	group~\cite{Wigner:1939cj}, and contrast it with the construction of
	carrollian fields of the previous subsection. For the massive UIR the
	isotropy subgroup is spanned by rotations and temporal and spatial
	translations which are generated by $\langle J_{ab}, H, P_a \rangle$.
	The inducing representation is then again given by
	\begin{align}
	\label{eq:JIR}
	\left[J_{ab},\phi^{\pm}_{a_1\ldots a_s}(\bzero)\right]&=\left(\Sigma_{ab}\right)^{b_1\ldots b_s}_{a_1\ldots a_s}\, \phi^{\pm}_{b_1\ldots b_s}(\bzero)\,, \qquad  \left[P_a,\phi^{\pm}(\bzero)\right]=0 \,, 
	\end{align}
	but since $H$ is now in the isotropy subgroup we also specify
	\begin{align}
	\label{eq:ind-massive}
	\left[H,\phi^{\pm}(\bzero)\right]=\pm i m\, \phi(\bzero)\,,
	\end{align}
	corresponding to a massive particle in the rest frame with momentum
	$p^{\pm}_{\mu} =\pm (m,\bzero)$. The induced representation is then obtained by boosting
	$\phi^{\pm}$ to a generic momentum frame,
	\begin{equation}
	\label{eq:Heisen-1}
	\phi^{\pm}(\x) = U(\x)\, \phi^{\pm}(\bzero)\, U(\x)^{-1}\,, 
	\end{equation}
	where we have defined the coordinates by $U: \HH^{d} \to G$ and
	$U(\x)=e^{x^{a}B_{a}}$.
	This implies
	\begin{align}
	\left[B_a,\phi^{\pm}(\bzero) \right]=\pd_{a}\phi^{\pm}(\bzero) \,,
	\end{align}
	and following similar computations as in
	Section~\ref{sec:carrollian-fields}, yields
	\begin{subequations}
		\label{eq:action-on-phi+}
		\begin{align}
		[J_{ab},\phi^\pm(\x)]&=-\xi_{J_{ab}}\phi^\pm(\x)+\Sigma_{ab}\, \phi^\pm(\x)\,, \\
		[B_a,\phi^\pm(\x)]&=-\xi_{B_{a}}\phi^\pm(\x)-\tanh(r/2) \hat{x}^b\Sigma_{ba}\, \phi^\pm(\x)\,,\\ 
		[H,\phi^\pm(\x)] &= \pm im \cosh(r)\, \phi^\pm(\x)\,, \label{eq:MH} \\
		[P_a,\phi^\pm(\x)]&=\mp im \sinh(r)\, \hat{x}_a\, \phi^\pm(\x)\,.\label{eq:MP}
		\end{align}
	\end{subequations}
	The main difference between the massive UIR and the $\zTi$ fields is
	the action of $H$ and $P_{a}$, which in contrast
	to~\eqref{eq:action-on-fields} acts by a phase and does not change the
	coordinates of the fields (cf.~Footnote~\ref{fn:1}). Indeed, their
	action yields the Lorentz boosted momentum
	$p^{\pm}_{\mu} = \pm ( m\cosh(r),- m\sinh(r) \hat x_a)$ in terms of
	rapidity $r$ and boost direction~$\hat x_{a}$ and provide us with the
	familiar equation
	\begin{align}
	\label{eq:p-eigen}
	[P_{\mu},\phi^{\pm}(\x)] = i p^{\pm}_{\mu} \phi^{\pm}(\x) \, .
	\end{align}
	Consequently, the action of the Casimir $\mathcal{C}_2=-H^2+P^a P_a$
	is given by
	\begin{align}
	\label{eq:casimir-fields-massive}
	\left[\mathcal{C}_2,\phi^{\pm}(\x)\right] = m^{2}\phi^{\pm}(\x)\,,
	\end{align}
	which ensures that the field is on the mass-shell. For fixed $m$ the
	Casimir is a multiple of the identity, as is necessary for an
	irreducible unitary representation. This can again be compared with
	our fields on $\zTi$ where we needed to impose the additional
	equation~\eqref{eq:reducibility} at the end of the construction.
	
	We are now in the position to compare the representation obtained in
	the last section to the classic Wigner construction recalled above. As
	is well-known, the representations $\phi^\pm(\x)$ defined above are
	both unitary and irreducible. On the other hand, the $\zTi$
	representation obtained in the last section is also unitary but
	reducible. Indeed, the latter can be inferred from Schur’s lemma for
	unitary representations and the fact that the Casimir
	\eqref{eq:casimir-fields} is not a multiple of the identity before
	\eqref{eq:reducibility} is imposed. More precisely, we will now argue
	that the $\zTi$ representation can be understood as a direct integral
	of irreducible representations; for a rigorous discussion of this
	concept see, e.g., Chapter 5 in \cite{Barut:1986dd}. From equations
	\eqref{eq:action-on-fields} and \eqref{eq:action-on-phi+} it is clear
	that the difference between the two representations lies only in the
	representation of $H$ and $P_a$. Let us therefore start with either
	one of the irreducible representations $\phi^\pm(\x)$ from which we
	construct
	\begin{equation}
	\label{eq:phi-relations}
	\phi^\pm_m(\x)=\frac{1}{2\pi}\int^\infty_{-\infty} \dd \tau \, e^{\mp im\tau} \phi(\tau,\x)\,, \qquad   \phi(\tau,\x)=\int^\infty_{-\infty} \dd m \,  e^{\pm im\tau}\phi^\pm_m(\x)\,,
	\end{equation}
	where we explicitly added the dependence
	of the Wigner representation on the mass $m$. A short
	computation shows that applying the transformations of
	\eqref{eq:action-on-fields} induces the Wigner transformations
	discussed above. We can therefore decompose a $\zTi$ field as a Fourier
	transform over irreducible representations. Explicitly we have
	\begin{align}
	[P^{\zTi}_\mu,\phi(\tau,\x)]=\int^\infty_{-\infty} \dd m \,e^{\pm im \tau} [P^{\mathrm{Wigner}}_\mu ,\phi^\pm_m(\x)] \, ,
	\end{align}
	where $P^{\zTi}_\mu$ denotes the action \eqref{eq:action-on-fields} on
	$\zTi$ fields and $P^{\mathrm{Wigner}}_\mu$ the action on the unitary
	irreducible representation \eqref{eq:action-on-phi+}.The reducible
	representation on the $\zTi$ fields is unitary with respect to the
	measure \eqref{eq:volumeform} on $\zTi$,
	\begin{equation}
	\label{eq:innerprod-ti}
	\langle \phi,\psi\rangle_{\zTi}=\frac{1}{2\pi} \int_{\zTi} \varepsilon\, \phi(\tau,\x) \cdot \psi(\tau,\x),
	\end{equation}
	where $\phi \cdot \psi$ is the pointwise inner product associated with
	the finite-dimensional $\SO(d)$ representation. Denoting the measure
	with respect to which the UIR \eqref{eq:action-on-phi+} is unitary by
	\begin{align}
	\label{eq:innerprod-UIR}
	\langle \phi,\psi \rangle_{m}=\int_{\HH}\varepsilon^{(d)}\phi^{*}_{m}(\x) \cdot  \psi_{m}(\x) 
	\end{align}
	where $m$ is the mass of the corresponding representation, the above
	relations allow to derive
	\begin{equation}
	\langle \phi,\psi\rangle_{\zTi}=\int^\infty_{-\infty} \dd m\, \langle \phi^{\pm},\psi^{\pm}\rangle_{m} \, .
	\end{equation}
	This demonstrates again clearly that the novel $\zTi$ representation
	discussed in the previous subsection is an integral over the
	well-known massive particle states.
	
	\section{Relations to fields in Minkowski space and massive celestial CFT}
	\label{sec:relat-fields-mink}
	
	In this section, we establish connections between the carrollian
	fields on $\zTi_{d+1}$ and massive fields living in Minkowski space
	$\MM_{d+1}$, as well as their formulation in terms of massive
	celestial conformal field theory.
	
	\subsection{Correspondence with massive fields in Minkowski space}
	\label{sec:connection-with-bulk}
	
	In the previous section we constructed massive carrollian field
	representations of $\ISO(d,1)$ living on $\zTi_{d+1}$. We will now
	show that these precisely arise when considering massive fields in
	$\mathbb{M}_{d+1}$ in the asymptotic limit to future/past timelike
	infinity. In particular it will be shown that massive carrollian
	fields encode massive particle UIRs in a natural way, and we
	will give their explicit decomposition in terms of creation and
	annihilation operators.
	
	We thus start with Minkowski spacetime $\mathbb{M}_{d+1}$ in cartesian
	coordinates $X^\mu$, in terms of which the Minkowski metric $\eta$ is given by
	\begin{equation}
	\label{eq:mink-met}
	\eta = \eta_{\mu\nu}dX^\mu dX^\nu\,,
	\end{equation}
	where $\eta_{\mu\nu} = \text{diag}(-1,1,\dots,1)$ denotes the
	components of the Minkowski metric in cartesian coordinates;
	throughout this section, the indices $\mu,\nu$ will be used to
	indicate components in cartesian coordinates. To describe the approach
	to the timelike infinities $i^\pm$ it is convenient to introduce a
	hyperbolic foliation of the inner lightcone emanating from the origin
	\begin{equation}
	\label{eq:hyperbolic-coordinates}
	X^\mu=\tau\, y^\mu(\x)\,, \qquad \eta_{\mu\nu}y^\mu y^\nu =-1\,.
	\end{equation}
	The spatial coordinates $\x=x^i$ for $i=1,\dots,d$ are defined
	in~\eqref{y to x} and provide an intrinsic chart of the quadric
	$\mathbb{H}^d$ characterised by the constraint $y^2=-1$, while the
	coordinate $\tau$ is identified with the proper distance from the
	origin $X^\mu=0$ to any point of the corresponding hyperbolic leaf of
	this foliation. Note that there is no difference between the indices
	$i,j$ or $a,b$ used in previous sections, namely they run over the
	same numbers $1,...,d$. However in this section will reserve $i,j$ for
	coordinate components and $a,b$ for orthonormal frame components to be
	defined momentarily. The coordinates $y^i$ and $x^i$ are formally
	identical to those we used in~\eqref{eq:invariants} and
	$\eqref{eq:invariants-2}$ to write down the carrollian structure on
	$\zTi_{d+1}$. Future and past timelike infinity $i^\pm$ are approached
	in the limit $\tau \to \pm \infty$, respectively. In the coordinate
	system $(\tau,\x)$ the Minkowski
	metric~\eqref{eq:mink-met} takes the form
	\begin{equation}
	\label{eq:hyperbolic-metric}
	\eta %= -d\tau^2 + \tau^2 \, h_{ij}(\y) \dd y^i \dd y^j
	= -d \tau^2+\tau^2\, h_{ij}(\x) \dd x^i \dd x^j\,,
	\end{equation}
	where $h_{ij}$ is the metric on $\HH^d$ that features
	in~\eqref{eq:invariants-2} and~\eqref{eq:invariants}. In cartesian
	coordinates the vector fields $\zeta_{X}$ with $X\in \iso(d,1)$ that
	generate Poincaré transformations are given by
	\begin{equation}
	\label{zeta}
	\zeta_{P_\mu}=\partial_\mu\,, \qquad \zeta_{J_{\mu\nu}}=X_\mu\, \partial_\nu - X_\nu\, \partial_\mu\,,
	\end{equation}
	where $X_\mu := \eta_{\mu\rho} X^\rho$. In order to express them in
	the hyperbolic coordinates $(\tau, \x)$, we introduce the projector
	$h_{\mu\nu}$ onto a single hyperboloid leaf of the foliation at
	fixed~$\tau$. In the ambient Minkowski space in cartesian coordinates,
	this hyperboloid is defined by
	\begin{equation}
	\label{eq:alternative-embedding}
	\eta_{\mu\nu}X^\mu X^\nu = -\tau^2\,,   
	\end{equation}
	where $\tau$ is constant, implying that the normal vector is
	$\tilde n_\mu = \eta_{\mu\nu}X^\nu$. Normalising the normal vector,
	i.e., requiring that $\eta^{\mu\nu}n_\mu n_\nu = -1$, tells us that
	\begin{equation}
	n_\mu = \tau^{-1}n_\mu = \tau^{-1}\eta_{\mu\nu} X^\nu = \eta_{\mu\nu}y^\nu =: y_\mu\,,    
	\end{equation}
	where we used~\eqref{eq:hyperbolic-coordinates}. Hence, the first
	fundamental form of a hyperboloid leaf embedded in $\MM_{d+1}$ is
	\begin{equation}
	\label{eq:projector}
	h_{\mu\nu} = \eta_{\mu\nu} + n_\mu n_\nu = \eta_{\mu\nu} + y_\mu y_\nu\,.
	\end{equation}
	On the other hand, the induced metric on the hyperboloid $\HH^d_\tau$
	at fixed $\tau = \text{const.}$ in~\eqref{eq:hyperbolic-metric} is
	$\eta\big\vert_{\HH^d_\tau} = \tau^2 h_{ij}(\x)\,dx^i\,dx^j$, which we
	can push forward to the first fundamental form~\eqref{eq:projector} as
	follows
	\begin{equation}
	h_{\mu\nu} := P^i_\mu\, h_{ij}\, P^j_\nu\,, \qquad  P^i_\mu := \tau\, \frac{\partial x^i}{\partial X^\mu}\,,
	\end{equation}
	where the components of $P^i_\mu$ are explicitly given by
	\begin{equation}
	\label{eq:projector-components-x}
	P^i_0=-\sinh r\, \hat x^i\,, \qquad P^i_j=\frac{r}{\sinh r}( \delta^i_j-\hat x^i \hat x_j)+\cosh r\, \hat x^i \hat x_j\,.
	\end{equation}
	Treating all Poincaré transformations \eqref{zeta} at once by using
	the standard parameterisation of an isometry of Minkowski space
	\begin{equation}
	\label{zeta mu}
	\zeta^\mu=A^\mu+\Omega^\mu{}_\nu\, X^\nu\,,
	\end{equation}
	with $A^\mu$ and $\Omega_{\mu\nu}=-\Omega_{\nu\mu}$ constants corresponding to
	translations and Lorentz rotations, we can then evaluate their
	components in $(\tau,\x)$ coordinates in the limit $\tau \to \infty$.
	This yields
	\begin{equation}
	\label{xi components}
	\begin{split}
	\xi^\tau&= \lim_{\tau \to \infty}\zeta^\tau=-y_\mu A^\mu=A^0 \cosh r-A^i \sinh r\, \hat x_i\,,\\
	\xi^i&= \lim_{\tau \to \infty} \zeta^i=P^i_\mu\, \Omega^{\mu\nu} y_\nu=\Omega^{ij} x_j-\left[r \coth r\,(\delta^i_j-\hat x^i \hat x_j)+\hat x^i \hat x_j \right]\, \Omega^{j0}\,,
	\end{split}
	\end{equation}
	where the last equality in each line is obtained by first writing out the expression in cartesian coordinates followed by the change of coordinates of~\eqref{y to x}; in particular, this means that the parameters $A^0$ etc.~that appear above are still the same constants that appear in~\eqref{zeta mu}. We recognise the expressions in~\eqref{xi components} as the components of the Killing vector
	fields on $\zTi_{d+1}$ given in \eqref{eq:fund-vector-fields}.
	Note also that the Casimir operator can be computed from
	$P_\mu A^\mu =\xi^\tau \partial_\tau$ with the choice
	that all the translations parameters $A^\mu$ are equal to one. This yields the expected result
	\begin{equation}
	\mathcal{C}_2=P^\mu P_\mu=-\partial_\tau^2\,.
	\end{equation}
	This provides the first hint that the carrollian structure of
	$\zTi_{d+1}$ arises from $\mathbb{M}_{d+1}$ in the limit
	$\tau \to \pm \infty$.
	
	We now turn to the study of massive fields in $\mathbb{M}_{d+1}$ and
	their behaviour in the limit $\tau \to \pm \infty$ where, as we shall
	demonstrate, carrollian fields on $\zTi_{d+1}$ emerge. In particular
	let us consider symmetric tensor fields
	$\phi_{\alpha_1\ldots \alpha_s}(X)$ carrying totally symmetric and
	traceless massive spin-$s$ particles. These satisfy the on-shell
	conditions \cite{Singh:1974qz,Singh:1974rc}
	\begin{subequations}
		\label{wave equation}
		\begin{align}
		\left(\Box-m^2 \right)\phi_{\alpha_1\ldots \alpha_s}&=0\,,\label{eq:wave-eq-1}\\
		\phi^\alpha{}_{\alpha\alpha_3 \ldots \alpha_s}=\nabla^\alpha \phi_{\alpha \alpha_2 \ldots  \alpha_s}&=0\,. \label{eq:wave-eq-2} 
		\end{align}
	\end{subequations}
	Although these conditions are those appropriate for the free fields
	that carry one-particle states, the analysis in fact also applies to
	interacting field theories which satisfy the cluster decomposition
	principle. Indeed, in that case the interactions are built
	perturbatively from the free theory, and appear in the Lagrangian
	through cubic and higher polynomials of the fields
	\cite{Weinberg:1995mt}. The resulting additional terms in the equation
	of motion do not alter the leading asymptotic behaviour of the fields
	in their limit towards $i^\pm$, which is that of the free fields.\footnote{Interactions with massless particles that mediate long-range forces change the asymptotic behaviour of massive fields in the limit $\tau \to \infty$; see, e.g., the analysis in \cite{Zwanziger:1973if,Campiglia:2015qka} for massive scalar QED. Such interactions are therefore excluded from our analysis.}
	
	In the coordinate system $x^\alpha=(\tau,x^i)$ given in \eqref{eq:hyperbolic-metric} the trace and
	divergence conditions~\eqref{eq:wave-eq-2} read
	\begin{equation}
	\label{trace and transversality}
	\eta^{ij} \phi_{ij \ldots }=\phi_{\tau \tau \ldots }\,, \qquad \eta^{ij} \nabla_i \phi_{j\ldots }=\nabla_\tau \phi_{\tau\ldots }\,,
	\end{equation}
	while the nonzero Christoffel symbols are given by
	\begin{equation}
	\label{eq:christoffel-asympt}
	\Gamma^\tau_{ij}=\tau^{-1}\, \eta_{ij}=\tau\, h_{ij}\,, \qquad \Gamma^i_{\tau j}=\tau^{-1} \delta^i_j\,, \qquad \Gamma_{ij}^k=\Gamma^k_{ij}[h]\,,
	\end{equation}
	where the $\Gamma^k_{ij}[h]$ are the components of the Levi-Civita
	connection on $\HH^d$ in $x^i$-coordinates. Using these, and
	introducing the shorthand notation
	\begin{equation}
	\phi_{i(k) \tau(s-k)} := \phi_{i_1\cdots i_k \!\!\underbrace{\tau\cdots \tau}_{(s-k) \, \text{times}}}\,,\qquad k=0,1,\ldots, s\,,
	\end{equation}
	a tedious but straightforward computation
	allows us to write the components of the wave operator acting on
	$\phi_{i(k) \tau(s-k)}$ in the form
	\begin{equation}
	\begin{split}
	\Box\, \phi_{i(k)\tau(s-k)}
	=&\left[-\partial_\tau^2-(d+2s-4k)\tau^{-1} \partial_\tau+\tau^{-2} (\triangle_{\HH}+c_0) \right]\phi_{i(k)\tau(s-k)}\\
	&-2k \tau^{-1} \nabla^{\HH}_{i(1)} \phi_{i(k-1)\tau(s-k+1)}+k(k-1) h_{i(2)} \phi_{i(k-2)\tau(s-k+2)}\,,
	\end{split}
	\end{equation}
	with $c_0\equiv k(d+2s+k)-(s-k)(d+s-k-1)$, and where the covariant
	derivative and laplacian on $\HH^d$ are denoted $\nabla^{\HH}_i$ and
	$\triangle_{\HH}=h^{ij} \nabla^{\HH}_i \nabla^{\HH}_j$, respectively.
	Note that the $k$ spatial indices are also implicitly symmetrised
	over.
	To simplify this equation we rescale the field components as
	\begin{equation}
	\label{rescaling}
	\phi_{i(k)\tau(s-k)}=\tau^{2k-s-\frac{d}{2}}\, \hat\phi_{i(k)\tau(s-k)}\,,
	\end{equation}
	such that the equation of motion \eqref{eq:wave-eq-1} reduces to
	\begin{equation}
	\label{wave equation components}
	\begin{split}
	0&=\left[-\partial_\tau^2+\tau^{-2} (\triangle_{\HH}+c_1)-m^2 \right]\hat \phi_{i(k)\tau(s-k)}\\
	&\quad -2k \tau^{-3} \nabla^{\HH}_{i(1)} \hat \phi_{i(k-1)\tau(s-k+1)}+k(k-1)\tau^{-4} h_{i(2)} \hat \phi_{i(k-2)\tau(s-k+2)}\,,
	\end{split}
	\end{equation}
	with $c_1=c_1(d,s,k)$ some constant which will not play any role in
	the subsequent analysis. From \eqref{wave equation components} we can
	infer that the asymptotic solution takes the form
	\begin{equation}
	\label{asymptotic solution}
	\hat \phi_{i(k)\tau(s-k)}(\tau,\x)=\phi_{i(k)\tau(s-k)}^+(\x)\, e^{im\tau}+\phi_{i(k)\tau(s-k)}^-(\x)\, e^{-im\tau}+\mathcal{O}(\tau^{-2})\, ,
	\end{equation}
	which is the solution to
	$(\pd^{2}_{\tau}+m^{2})\,\hat \phi_{i(k)\tau(s-k)}=0$ and which we can
	already recognise as the Casimir equation~\eqref{eq:reducibility}. The
	spatial components $\phi^\pm_{i(s)}(\x)$ are completely unconstrained
	while the divergence-free condition in \eqref{trace and
		transversality} completely determines the remaining time components
	$\phi^\pm_{i(k)\tau(s-k)}(\x)$ for $k < s$. Indeed
	$\phi^\pm_{i(s)}(\x)=\phi^\pm_{i_1\ldots i_s}(\x)$ carry the physical
	degrees of freedom of the corresponding spin-$s$ representation of the
	little group $\SO(d)$. The subleading components of the field can be
	determined recursively from this free asymptotic data. This together
	with \eqref{eq:solution} suggests that the spatial components of
	\eqref{asymptotic solution} provide a concrete realisation of the
	massive carrollian fields constructed in
	Section~\ref{sec:massive-reps-and-fields-at-Ti}. This will be
	confirmed by studying their Poincaré transformations.
	
	Let us therefore look at the transformation of the physical asymptotic
	degrees of freedom
	\begin{equation}
	\label{induced carrollian field}
	\bar \phi_{i_1\ldots i_s}(\tau,\x) := \phi_{i_1\ldots i_s}^+(\x)\, e^{im\tau}+\phi_{i_1\ldots i_s}^-(\x)\, e^{-im\tau}\,,
	\end{equation}
	and demonstrate that their transformation under Poincaré symmetries is
	exactly that of the massive carrollian fields on $\zTi_{d+1}$. First
	note that the transformation laws obtained in
	Section~\ref{sec:carrollian-fields} are appropriate for for the
	tangent space components of the fields, which carry indices
	$a_1,\dots,a_s$ as in~\eqref{induced carrollian field}, while the
	discussion of fields in Minkowski space so far applied to the
	coordinate components in hyperbolic coordinates. Locally flat
	coordinate systems can be reached with the use of vielbeins
	$E^A_\alpha$ such that
	\begin{equation}
	\eta_{\alpha\beta}=E^A_\alpha\, \eta_{AB}\, E^B_\beta\,,
	\end{equation}
	where $\eta_{\alpha\beta}$ represents the metric components in
	arbitrary curvilinear coordinates $x^\alpha$, while
	$\eta_{AB}=\text{diag}(-1,1,\ldots ,1)$ is the tangent space Minkowski
	metric. The tangent space indices $A,B = 0,\dots,d$ split according to
	$A = (0,a)$, where $a$ the spatial tangent space index. For
	$x^\alpha=(\tau,x^i)$ related to cartesian coordinates as
	in~\eqref{eq:hyperbolic-coordinates}, we can write the vielbeins that
	make up the Minkowski metric~\eqref{eq:hyperbolic-metric} as
	\begin{equation}
	E^A_\alpha=\begin{pmatrix}
	1 & 0\\
	0 & \tau e^a_i
	\end{pmatrix}\,, \qquad E_A^\alpha=\begin{pmatrix}
	1 & 0\\
	0 & \tau^{-1} e^i_a
	\end{pmatrix}\,,
	\end{equation}
	where the reduced vielbein $e^a_i$ and its inverse $e^i_a$ are respectively given by
	\begin{equation}
	e^a_i=\frac{\sinh r}{r} \delta^a_i+\left(1-\frac{\sinh r}{r} \right) \hat x^a \hat x_i\,, \quad e^i_a=\frac{r}{\sinh r} \delta^i_a+\left(1-\frac{r}{\sinh r}\right)\hat x^i \hat x_a\,,
	\end{equation}
	such that 
	\begin{equation}
	h_{ij}=e^a_i\, \delta_{ab}\, e^b_j\,,
	\end{equation}
	where $h_{ij}$ is the hyperbolic metric defined
	in~\eqref{eq:invariants}. The tangent space components of the
	asymptotic fields are then identified as
	\begin{equation}
	\label{coordinate to tangent}
	\bar \phi_{a_1 \ldots a_s}:=e_{a_1}^{i_1}\ldots e_{a_s}^{i_s}\, \bar \phi_{i_1\ldots i_s}\,.
	\end{equation}
	The Poincaré transformation of the bulk components
	$\phi_{A_1\ldots A_s}$ can be written
	\begin{equation}
	\delta_\zeta \phi_{A_1\ldots A_s}=-\zeta^\alpha \partial_\alpha \phi_{A_1\ldots A_s}+\Lambda\indices{_{A_1}^{B}} \phi_{B\ldots A_s}+\ldots +\Lambda\indices{_{A_s}^{B}} \phi_{A_1\ldots B}\,,
	\end{equation}
	in terms of the Killing vector field \eqref{zeta mu}, and with the
	local Lorentz transformation $\Lambda_{AB}$ chosen such as to leave
	the background tetrad $E^\alpha_A$ invariant,
	\begin{equation}
	\Lambda\indices{_A^B}=E^{B}_\alpha \Ld_\zeta E^\alpha_A=E^{B}_\alpha \left(\zeta^\beta \partial_\beta E^\alpha_A+E^\beta_A \partial_\beta \zeta^\alpha \right)\,.
	\end{equation}
	Using this and the falloff behavior \eqref{rescaling} we obtain the induced transformation of the asymptotic
	field components
	\begin{equation}
	\begin{split}
	\delta \bar \phi_{a_1\ldots a_s}&=-(\xi^\tau \partial_\tau+\xi^i \partial_i)\bar \phi_{a_1\ldots a_s}+\Lambda\indices{_{a_1}^{b}}\, \bar \phi_{b\ldots a_s}+\ldots +\Lambda\indices{_{a_s}^{b}}\, \bar \phi_{a_1\ldots b}\,,
	\end{split}
	\end{equation}
	with 
	\begin{equation}
	\label{LLT}
	\Lambda\indices{_a^b}=e^b_i\, (\xi^j \partial_j e^i_a+e^j_a \partial_j \xi^i )=\Omega\indices{_a^b}+\tanh(r/2)\, (\hat x_a \delta^b_c-\hat x^b \delta_{ac}) \Omega^{0c}\,,
	\end{equation}
	expressed solely in terms of the vector fields \eqref{xi components}
	on $\zTi_{d+1}$, and in agreement with the transformations of the fields
	constructed in Section~\ref{sec:carrollian-fields} and given in
	\eqref{eq:action-on-fields}. For completeness we also note that the
	transformation of the coordinate components can be fully covariantised
	on $\zTi_{d+1}$ by adding trivial time-components to the tensor
	$\bar \phi$, namely
	\begin{equation}
	n^\alpha \bar \phi_{\alpha\ldots \alpha_s}=0\,, \qquad \delta \bar \phi_{\alpha_1\ldots \alpha_s}=-\Ld_\xi\, \bar \phi_{\alpha_1\ldots \alpha_s}\,,
	\end{equation}
	where $n=\partial_\tau$ is the carrollian vector field of \eqref{eq:invariants}.
	
	We can summarise the correspondence we just established between
	massive fields $\phi_{\alpha_1 \ldots \alpha_s}$ on $\mathbb{M}_{d+1}$
	and carrollian fields $\bar \phi_{a_1 \ldots a_s}$ on $\zTi_{d+1}$
	through the following \textit{extrapolate dictionary}
	\begin{equation}
	\label{extrapolate dictionary}
	\bar \phi_{a_1\ldots a_s}\overset{\tau \to \infty}{\sim} \tau^{\frac{d}{2}} E_{a_1}^{\alpha_1} \ldots E_{a_s}^{\alpha_s} \phi_{\alpha_1 \ldots \alpha_s} \,, 
	\end{equation}
	where the limit is asymptotic in the sense that the carrollian field
	$\bar \phi$ still depends on $\tau$ through the phase factors
	$e^{\pm i m \tau}$.
	
	\subsection{Plane wave basis and reconstruction algorithm}
	\label{sec:relation-plane-waves}
	
	It is useful and informative to relate the above considerations to the
	standard plane wave decomposition of the massive field in cartesian
	coordinates $X^\mu$
	\begin{equation}
	\label{plane wave expansion}
	\phi_{\mu_1\ldots \mu_s}(X)=\sum_\sigma \int \frac{d^d\vec{k}}{2k^0 (2\pi)^d}\left(\epsilon^\sigma_{\mu_1\ldots \mu_s}(k)\, a^\sigma_k\, e^{ik \cdot X}+\epsilon^\sigma_{\mu_1\ldots \mu_s}(k)^* \,a^{\sigma\,\dagger}_k\, e^{-i k \cdot  X}\right)\,,
	\end{equation}
	where $k^{0}=\sqrt{m^{2}+\vec k^{2}}$. The sum is over all independent
	polarisations $\sigma$ and where the polarisation tensor
	$\epsilon^\sigma_{\mu_1\ldots \mu_s}$ is totally symmetric and
	traceless, and further satisfies the transversality condition
	$k^\mu \epsilon^\sigma_{\mu\ldots \mu_s}(k)=0$, cf.~\eqref{wave
		equation}. The modes $a^\sigma_k$ obey the canonical commutation
	relations
	\begin{equation}
	\label{eq:poissonbracketsa}
	[a^\sigma_k,a^{\sigma'\dagger}_{k'}]=2k^0(2\pi)^d\delta^{(d)}(\vec{k}-\vec{k}') \delta^{\sigma,\sigma'}.
	\end{equation}
	We will evaluate \eqref{plane wave expansion} in the limit
	$\tau \to \infty$ in order to extract an expression for the carrollian
	field \eqref{induced carrollian field}. In order to achieve this we
	once more turn to the hyperbolic coordinate system defined via $X^\mu=\tau y^\mu$
	with (cf.~\eqref{y to x})
	\begin{equation}
	\label{eq:hyppar}
	y^\mu=\left(\sqrt{1+\y\cdot \y},\y\right)\,, \qquad y^2=-1\,,
	\end{equation}
	and we similarly parameterise the massive momentum as
	$k^\mu=m\, \hat{k}^\mu$ through
	\begin{equation}
	\label{k hyppar}
	\hat{k}^\mu=\left(\sqrt{1+\k\cdot \k},\k\right)\,, \qquad  \hat{k}^2=-1\,,
	\end{equation}
	such that \eqref{plane wave expansion} can be rewritten as
	\begin{equation}
	\label{eq:16}
	\phi_{\mu_1\ldots \mu_s}(X)=\frac{m^{d-1}}{2(2\pi)^d} \sum_\sigma \int \frac{d^d\k}{\sqrt{1+\k\cdot \k}}\left(\epsilon^\sigma_{\mu_1\ldots \mu_s}(k)\, a^\sigma_k\, e^{im\tau \hat k \cdot y}+ \text{h.c.}\right)\,.
	\end{equation}
	The asymptotic limit $\tau \to \infty$ can be evaluated by stationary
	phase approximation as done for instance in \cite{Campiglia:2015qka}.
	In the parametrisation \eqref{eq:hyppar}--\eqref{k hyppar} we have
	\begin{equation}
	\label{eq:17}
	\hat{k} \cdot y=-\sqrt{1+\k\cdot \k}\, \sqrt{1+\y\cdot \y}+\k \cdot \y\,,
	\end{equation}
	such that the critical point of \eqref{eq:16} at large $\tau$ is located at
	\begin{equation}
	\label{eq:18}
	\k=\y\,.
	\end{equation}
	The determinant of the second derivative of \eqref{eq:17} at the
	critical point is given by
	\begin{equation}
	\label{eq:19}
	\begin{split}
	\det\left(m \partial_{k^i}\partial_{k^j}( \hat{k} \cdot y)\right)
	&=\det\left(m\frac{\sqrt{1+\y^2}}{\sqrt{1+\k^2}}\left(-\delta_{ij}+\frac{k_ik_j}{1+\k^2}\right)\right)\\
	&\overset{\y=\k}{=}\det\left(m\left(-\delta_{ij}+\frac{y_iy_j}{1+\y^2}\right)\right)
	=\frac{(-m)^d}{1+\y^2}
	=\frac{(-m)^d}{(y^0)^{2}}\,.    
	\end{split}
	\end{equation}
	Taken together, the stationary phase approximation of \eqref{eq:16}
	yields
	\begin{equation}
	\label{eq:sattle-result}
	\phi_{\mu_1\ldots \mu_s}(X)\overset{\tau \to \infty}{\approx}\frac{m^{d/2-1}}{2(2\pi \tau)^{d/2}} \sum_\sigma \left(\epsilon^\sigma_{\mu_1\ldots \mu_s}(k)\, a^\sigma_k\, e^{-im\tau}+\text{h.c.}\right)\Big|_{k=m y}\,.
	\end{equation}
	We are actually interested in the spatial components in hyperbolic
	coordinates $(\tau, \y)$, obtained from the cartesian components given
	above through the jacobian transformation\footnote{Using
		\eqref{eq:sattle-result} it would seem that
		$\phi_{i(k)\tau(s-k)}=\mathcal{O}(\tau^{k-\frac{d}{2}})$ instead of
		$\mathcal{O}(\tau^{2k-s-\frac{d}{2}})$ as predicted in
		\eqref{rescaling}. However note that the transversality condition
		$ k^\mu \epsilon_{\mu\ldots\mu_s}(k)=0$ makes this seemingly
		overleading contribution disappear due to the critical
		point $\hat k=y$.}
	\begin{equation}
	\phi_{i(k)\tau(s-k)}=\tau^k\, \partial_{i_1} y^{\mu_1}\ldots \partial_{i_k} y^{\mu_k}\, y^{\mu_{k+1}} \ldots y^{\mu_s}\,  \phi_{\mu_1\ldots \mu_s}\,.
	\end{equation}
	The degrees of freedom of the massive carrollian field sit in the purely spatial components, thus given by
	\begin{equation}
	\label{asymptotic phi_a}
	\begin{split}
	\phi_{i_1\ldots i_s}(X)&=\tau^s\, \partial_{i_1} y^{\mu_1}\ldots \partial_{i_s} y^{\mu_s}\, \phi_{\mu_1\ldots \mu_s}\\
	&\overset{\tau \to \infty}{\approx} \frac{m^{d/2-1}}{2(2\pi)^{d/2}}\, \tau^{s-\frac{d}{2}} \sum_\sigma \left(\epsilon^\sigma_{i_1\ldots i_s}(k)\, a^\sigma_k\, e^{-im\tau}+\text{h.c.}\right)\Big|_{k=m y}\,,
	\end{split}
	\end{equation}
	where we have defined the spatial polarisations
	\begin{equation}
	\epsilon^\sigma_{i_1\ldots i_s}(k)\equiv \partial_{i_1} y^{\mu_1}\ldots \partial_{i_s} y^{\mu_s}\, \epsilon^\sigma_{\mu_1\ldots \mu_s}(k)\big|_{y=\hat k}\,.
	\end{equation}
	Note that there are exactly as many independent polarisations $\sigma$
	as there are independent totally symmetric and traceless tensors
	$\epsilon_{i_1\ldots i_s}$, which is just another way to say that
	$\epsilon^\sigma_{i_1\ldots i_s}$ carries the spin-$s$ irrep of
	$\SO(d)$. The leading $\tau^{s-\frac{d}{2}}$ behavior of
	\eqref{asymptotic phi_a} is indeed the one anticipated in
	\eqref{rescaling} for $k=s$, and comparison with \eqref{asymptotic
		solution} allows us to identify the carrollian degrees of freedom
	\begin{equation}
	\label{phi=a}
	\begin{split}
	\phi^{+}_{i_{1}\ldots i_{s}}(\y)&= \frac{m^{d/2-1}}{2(2\pi)^{d/2}} \sum_\sigma  \epsilon^\sigma_{i_1\ldots i_s}(k)\, a^\sigma_k\, \Big|_{k=m y(\y)}\,,\\
	\phi^{-}_{i_{1}\ldots i_{s}}(\y)&=\frac{m^{d/2-1}}{2(2\pi)^{d/2}} \sum_\sigma \epsilon^{\sigma}_{i_1\ldots i_s}(k)^*\, a_k^{\sigma\, \dagger} \Big|_{k=m y(\y)}\,. 
	\end{split}
	\end{equation}
	This result makes fully explicit the previous statement that
	\textit{the massive carrollian fields essentially are the particle
		degrees of freedom}, up to the change of basis given in
	\eqref{coordinate to tangent}. On the one hand the asymptotic leaf
	$\HH^d$ of the hyperbolic foliation \eqref{eq:hyperbolic-metric} is
	explicitly identified with the massive momentum shell through the
	critical point $y=\hat k$, while the spin-$s$ representation
	of the little group $\SO(d)$ is manifestly carried by the spatial
	polarisation tensor $\epsilon^\sigma_{i_1\ldots i_s}$.
	
	We will close this section by noting that the relations \eqref{phi=a},
	when inverted such as to express $a^\sigma_k$ in terms of the
	components of $\phi^+_{i_1\ldots i_s}$ and upon replacement back into
	the original plane wave expansion \eqref{plane wave expansion},
	provide a reconstruction algorithm for the relativistic bulk fields
	$\phi_{\mu_1\ldots \mu_s}(X)$ from its asymptotic value
	$\phi_{i_1\ldots i_s}(\y)$ near timelike infinity. Explicitly for the
	scalar field this straightforwardly yields
	\begin{equation}
	\phi(X)=m^{1-\frac{d}{2}} \int \frac{d^d\vec{k}}{k^0(2\pi)^{d/2}}\left( \phi^{+}(\k)\, e^{ik \cdot X}+\phi^{-}(\k)\, e^{-ik \cdot X}\right)\,.
	\end{equation}
	This is the timelike analog of the Kirchhoff--d’Adhémar
	formula~\cite{Penrose:1980yx,Penrose:1987uia}. From a carrollian
	perspective this is the $\zTi$ analog of what has been considered for null
	infinity in Section~2.2.2 of~\cite{Donnay:2022wvx}.
	
	\subsection{Relation to massive conformal primary wavefunctions}
	\label{sec:relat-mass-celest}
	
	Having established the relation between plane wave creation and
	annihilation operators and the fields $\phi_\pm(\x)$ transforming
	under a representation of $\zTi$ at fixed Casimir, we can straightforwardly relate the
	latter to celestial CFT variables. In the following we rely on the
	discussion in \cite{Pasterski:2017kqt} and restrict ourselves to the
	case of a scalar field for simplicity.
	
	In the celestial CFT context, one considers the following
	\emph{massive conformal primary} basis of solutions to the free wave
	equation
	\begin{equation}
	\label{eq:22}
	\phi^\pm_\Delta(X^\mu;\vec{w})=\int_{\HH^{d}}\frac{d^{d}\hat{k}}{\hat{k}^0} G_{\Delta}(\hat{k};\vec{w})e^{\pm i m \hat{k} \cdot X}.
	\end{equation}
	Here, $\vec{w}\in \RR^{d-1}$ is a point on the boundary of $\HH^{d}$ and $G_{\Delta}(\hat{k};\vec{w})$ is the bulk-to-boundary propagator on $\HH^{d}$ that can be written as 
	\begin{equation}
	\label{eq:23}
	G_{\Delta}(\hat{k};\vec{w})=\frac{1}{(-\hat{k}\cdot q)^\Delta}\,,
	\end{equation}
	where
	\begin{equation}
	\label{eq:24}
	q^\mu(\vec{w})=(1+\vec{w}\cdot \vec{w},2\vec{w},1-\vec{w}\cdot \vec{w})\,.
	\end{equation}
	Note that in \eqref{eq:22} one trades the $d$ spatial momentum
	components specifying a plane wave solution of given mass for a weight
	$\Delta$ and $d-1$ points on the boundary of the hyperboloid, thus
	staying with the same number of data. It can be shown
	that one needs to set $\Delta=d/2+i \nu$ with
	$\nu\in \mathbb{R}$ in order to have delta-function normalisable
	modes~\cite{Pasterski:2017kqt}.
	
	Instead of expanding a solution to the massive Klein--Gordon equation
	in terms of plane waves and corresponding creation and annihilation
	operators $a_k,a^\dagger_k$ as in \eqref{plane wave expansion}, one
	expands the field into conformal primary waves
	$\phi^{\pm}_{\Delta}(X^\mu;\vec{w})$ and \emph{celestial operators}
	\begin{equation}
	\label{eq:celO}
	\mathcal{O}^{\Delta}(\vec{w})=m^{d-1}\int \frac{d^d\hat{k}}{\hat{k}^{0}(2\pi)^d}a_{m\hat{k}}\,G_\Delta(\hat{k};\vec{w})\,
	\end{equation}
	with inverse relation
	\begin{equation}
	\label{eq:celinv}
	a_{m\hat{k}}=m^{1-d}(2\pi)^d\int^{\infty}_{-\infty}\dd \nu \mu(\nu)\int \dd \vec{w}\,G_{\Delta^*}(\hat{k};\vec{w})\mathcal{O}^{\Delta}(\vec{w}),
	\end{equation}
	where $\mu(\nu)$ is a normalisation factor \cite{Costa:2014kfa, Pasterski:2017kqt}.
	
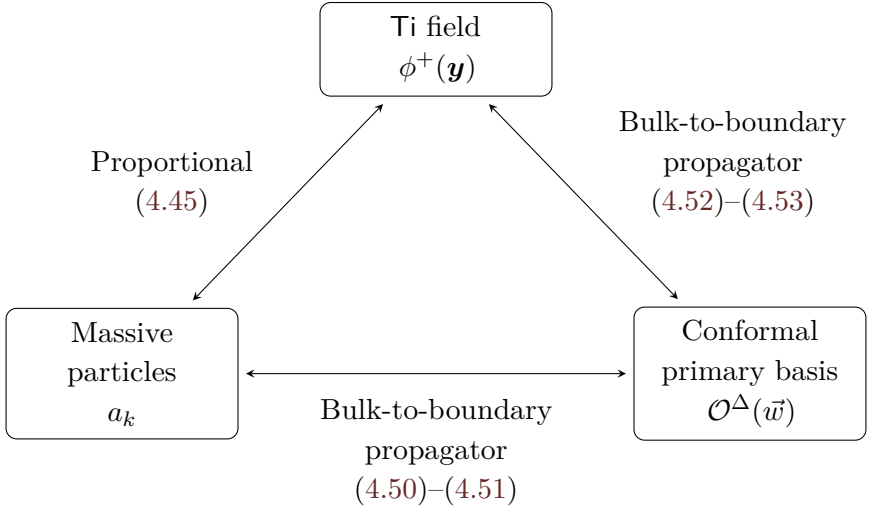
\begin{figure}[ht!]
		\centering
		\begin{tikzpicture}
		\tikzstyle{nb} = [rectangle,draw,fill=white,rounded corners,outer sep=3pt,text width=2.7cm,align=center,inner sep=5pt];
		\coordinate (A) at (-2.5, 0);
		\coordinate (B) at ( 2.5, 0);
		\coordinate (C) at ( 0, 4.3);
		\node[nb,left] (An) at (A) {Massive particles\\ $a_k$};
		\node[nb,right] (Bn) at (B) {Conformal primary basis\\ $\mathcal{O}^{\Delta}(\vec{w})$};
		\node[nb] (Cn) at (C) {$\zTi$ field\\ $\phi^{+}(\y)$};
		\draw[stealth-stealth] (An) -- (Bn);
		\draw[stealth-stealth] (Bn) -- (Cn);
		\draw[stealth-stealth] (An) -- (Cn);
		\node[below,align=center,text width=4cm] at ($(A)!0.5!(B)-(0,0.2)$) {Bulk-to-boundary propagator\\
			\eqref{eq:celO}--\eqref{eq:celinv}};
		\node[align=center,anchor=south east] at ($(An)!0.45!(Cn)$) {Proportional \\ \eqref{phi=a}};
		\node[align=center,anchor=south west] at ($(Bn)!0.45!(Cn)$) {Bulk-to-boundary\\propagator\\ \eqref{eq:ccfttoti}--\eqref{eq:titoccft}};
		\end{tikzpicture}
		\caption{Three different perspectives on massive states in
			flat spacetime and their interrelations: carrollian (top),
			momentum space (left) and celestial (right).}
		\label{fig:triangle}
	\end{figure}

	Since the fields on $\zTi_{d+1}$ are proportional to the plane wave
	creation and annihilation operators, the relation is now obvious.
	Explicitly, we find from~\eqref{phi=a}
	\begin{equation}
	\label{eq:ccfttoti}
	\mathcal{O}^{\Delta}(\vec{w})=2m^{d/2}\int \frac{d^d\y}{\hat{X}^0(\y)(2\pi)^{d/2}}\phi^+(\y)G_\Delta(\hat{X}(\y);\vec{w}),
	\end{equation}
	and
	\begin{equation}
	\label{eq:titoccft}
	\phi^+(\y)= \frac{m^{-d/2}}{2}(2\pi)^{d/2}\int^{\infty}_{-\infty}\dd \nu \mu(\nu)\int \dd \vec{w}\,G_{\Delta^*}(\hat{X}(\y);\vec{w})\,\mathcal{O}^{\Delta}(\vec{w}).
	\end{equation}
	We have summarised these interrelations between the momentum, carrollian
	and celestial points of view in Figure~\ref{fig:triangle}, which can be
	contrasted with the analogue for null infinity in~\cite{Donnay:2022wvx}.
	
	\section{Carrollian energy-momentum tensor and conservation laws}
	\label{sec:EMT-and-conservation-laws}
	
	In this section, we provide an intrinsic definition of a carrollian
	energy-momentum tensor (EMT) on $\zTi_{d+1}$. We then explicitly
	construct the EMT of a massive scalar in two complementary ways: first
	by pushing the lorentzian EMT from Minkowski to timelike infinity
	using the techniques of Section~\ref{sec:connection-with-bulk}, and
	then by varying the lagrangian of an electric Carroll scalar field
	theory. Finally, we discuss the relation between additional conserved
	charges corresponding to the $\zTi$-supertranslations~\eqref{eq:H-bms}
	and BMS charges as well as their connection to soft graviton theorems.

	\subsection{Intrinsic description}
	\label{sec:EMT-intrinsic}
	
	We begin with a discussion carrollian EMTs from a variational
	perspective (see,
	e.g.,~\cite{deBoer:2017ing,deBoer:2020xlc,Baiguera:2022lsw,deBoer:2023fnj,Armas:2023dcz}, which allows us to derive the general properties of carrollian EMTs.
	At the same time, this procedure provides an explicit recipe for
	computing the EMT of a given theory by coupling it to an arbitrary
	carrollian background.
	
	We recall that in a relativistic theory, the EMT may be extracted by
	coupling the theory to a curved lorentzian background and varying with
	respect to the metric, and further arises as the conserved current
	associated with diffeomorphism symmetry. In other words, given a
	relativistic field theory coupled to a lorentzian background with
	metric $g_{\mu\nu}$ described by an action
	$S_{\text{R}}[\varphi_I;g_{\mu\nu}]$, where $\varphi_I$ abstractly
	captures the field content, the variation takes the form
	\begin{equation}
	\delta S_{\text{R}}[\varphi_I;g_{\mu\nu}] = \int d^{d+1}x\,\sqrt{-g}\left[  \frac{1}{2}T^{\mu\nu}_{\text{R}}\delta g_{\mu\nu} + \mathcal{E}_{\text{R}}^I\delta\varphi_I  \right]\,,
	\end{equation}
	where $T^{\mu\nu}_{\text{R}}$ is the relativistic EMT, and
	$\mathcal{E}_{\text{R}}^I$ are the equations of motion for the fields
	$\varphi_I$. The symmetry of $g_{\mu\nu}$ immediately tells us that
	$T^{\mu\nu}_{\text{R}}$ is symmetric, while diffeomorphism invariance
	of the action leads to conservation of $T^{\mu\nu}_{\text{R}}$ when
	$\mathcal{E}_{\text{R}}^I = 0$. If we denote the Levi-Civita
	connection of $g_{\mu\nu}$ by $\hat\nabla$, the metric transforms as
	$\delta_\zeta g_{\mu\nu} = \pounds_\zeta g_{\mu\nu} =
	2\hat\nabla_{(\mu}\zeta_{\nu)}$ under an infinitesimal diffeomorphism
	generated by $\zeta^\mu$, and on-shell invariance of $S_{\text{R}}$
	implies that
	\begin{equation}
	\label{eq:rel-diffeo-calc}
	0 = \delta_\zeta S_{\text{R}}[\varphi_I;g_{\mu\nu}] = \int d^{d+1}x\,\sqrt{-g}\,T^{\mu\nu}_{\text{R}}\hat\nabla_{\mu}\zeta_{\nu} = - \int d^{d+1}x\,\sqrt{-g}\,\zeta_{\nu}\hat\nabla_\mu T^{\mu\nu}_{\text{R}} \,,
	\end{equation}
	where we used symmetry of $T^{\mu\nu}_{\text{R}}$ and threw away
	boundary terms arising from integration by parts. Since the above must
	vanish for arbitrary $\zeta^\mu$, we must have that
	\begin{equation}
	\hat\nabla_\mu T^{\mu\nu}_{\text{R}} = 0\,.
	\end{equation}
	This is the diffeomorphism Ward identity for a relativistic field theory.
	
	For a carrollian theory with generic field content $\Phi_I$ for $I$
	some abstract label, the same principle applies: coupling it to an
	arbitrary carrollian geometry allows us to extract the carrollian EMT.
	More precisely, if the carrollian background is given in terms of the
	carrollian structure $(n^\mu,q_{\mu\nu})$ with inverses
	$(\tau_\mu,\gamma^{\mu\nu})$ the variation of the action
	$S[\Phi_I;\tau_\mu,q_{\mu\nu}]$ can be written
	as~\cite{deBoer:2017ing,deBoer:2023fnj,Armas:2023dcz,Baiguera:2022lsw}
	\begin{equation}
	\delta S[ \Phi_I;\tau_\mu,q_{\mu\nu} ] = \int d^{d+1}x\, e\left[ T^\mu \delta \tau_\mu + \frac{1}{2}\mathcal{T}^{\mu\nu}\delta q_{\mu\nu} + \mathcal{E}^I\delta\Phi_I \right]\,,
	\end{equation}
	where the measure\footnote{For $\zTi_{d+1}$, the integral measure
		$e\,d^{d+1} x = \varepsilon$ is the invariant volume form defined
		in~\eqref{eq:volumeform}.} is
	$e = \sqrt{\det(\tau_\mu\tau_\nu + q_{\mu\nu})}$, and where $T^\mu$ is
	the energy current, $\mathcal{T}^{\mu\nu} = \mathcal{T}^{\nu\mu}$ is
	the momentum-stress tensor, and $\mathcal{E}^I$ represents the
	equations of motion for the fields $\Phi_I$. Computing the second
	variation under local carroll boosts, $\delta_C(\delta_C S) = 0$, and
	assuming that $\Phi_I$ is Carroll boost invariant, the fact that
	$\tau_\mu$ transforms as in~\eqref{eq:boost-trafos} implies that
	$\mathcal{T}^{\mu\nu}$ is not boost invariant:
	$\delta_C \mathcal{T}^{\mu\nu}=
	-2T^{(\mu}\gamma^{\nu)\rho}\lambda_\rho$, while the energy current
	$T^\mu$ remains invariant (see~\cite{Armas:2023dcz} for more details).
	Furthermore, requiring that $S[ \Phi_I;\tau_\mu,q_{\mu\nu} ]$ is
	invariant under local Carroll boosts~\eqref{eq:boost-trafos} yields
	the boost Ward identity
	\begin{equation}
	\label{eq:boost-WI}
	T^\mu q_{\mu\nu} = 0\,. 
	\end{equation}
	This relation is true off-shell\footnote{This statement is only true
		because we assumed that the $\Phi^I$ are inert under Carroll boosts.
		If they were not, the equations of motion $\mathcal{E}^I$ would
		feature on the right-hand side of~\eqref{eq:boost-WI}.}, which means
	that we may without loss of generality express the energy current as
	\begin{equation}
	T^\mu = n^\mu {E}\,, \qquad E \equiv \tau_\mu T^\mu\,,
	\end{equation}
	where we used~\eqref{eq:carroll-rels}. The quantity $E$ is a boost
	invariant scalar which we can think of as an energy. Finally, when
	$\mathcal{E}^I = 0$, i.e., when the $\Phi_I$ are on-shell, a
	calculation analogous to~\eqref{eq:rel-diffeo-calc} shows that
	diffeomorphism invariance gives rise to the `conservation' equation
	\begin{equation}
	\label{eq:diffeo-WI-torsion}
	e^{-1}\D_\nu (e\,T^\nu{_\mu}) = T^\nu\D_\mu \tau_\nu + \frac{1}{2}\mathcal{T}^{\nu\rho}\D_\mu q_{\nu\rho}\,,
	\end{equation}
	where we defined the EMT
	\begin{equation}
	T^\mu{_\nu} = \tau_\nu T^\mu + \mathcal{T}^{\mu\rho}q_{\rho\nu}\,,
	\end{equation}
	which is boost invariant by virtue of the boost Ward
	identity~\eqref{eq:boost-WI}. This means that, given a carrollian
	Killing vector $\xi^\mu$ satisfying~\eqref{eq:carrollian-KVF}, the
	contraction of~\eqref{eq:diffeo-WI-torsion} with $\xi^\mu$ gives
	\begin{equation}
	\label{eq:general-conservation}
	0 = e^{-1}\D_\nu (e\,T^\nu{_\mu} \xi^\mu) + E\tau_\mu \pounds_\xi n^\mu - \frac{1}{2} \mathcal{T}^{\mu\nu}\pounds_\xi q_{\mu\nu} =  e^{-1}\D_\nu (e\,T^\nu{_\mu} \xi^\mu)\,,
	\end{equation}
	where we used~\eqref{eq:carrollian-KVF}. Hence, the combination $j^\mu = T^\mu{_\nu} \xi^\nu$ is a conserved current. 
	
	In what follows, we shall assume the existence of an adapted and
	torsion-free carrollian connection $\nabla$ as discussed around
	Eq.~\eqref{eq:torsion-free-connection}, in terms of which the
	diffeomorphism Ward identity~\eqref{eq:diffeo-WI-torsion} becomes
	\begin{equation}
	\label{eq:diffeo-WI}
	\nabla_\nu T^\nu{_\mu} = 0\,,
	\end{equation}
	while the current conservation~\eqref{eq:general-conservation} takes the form
	\begin{equation}
	\label{eq:conservation-torsion-free}
	\nabla_\mu (T^\mu{_\nu}\xi^\nu) = 0\,,
	\end{equation}
	which is identical in form to the corresponding relation in lorentzian
	theories.
	
	To define the EMT on $\zTi_{d+1}$ and derive the currents for a given
	theory, we therefore should couple the theory to a generic carrollian
	geometry and vary the background as detailed above, and then
	specialise to the case of $\zTi_{d+1}$ once the general expressions
	have been found. Alternatively, we may construct the conservation laws
	directly on $\zTi_{d+1}$ using its carrollian
	structure~\eqref{eq:invariants}.
	
	Turning~\eqref{eq:conservation-torsion-free} on its head, we may
	\textit{define} the EMT $T\indices{^{\mu}_{\nu}}$ on $\zTi_{d+1}$ by
	demanding that its contraction with a Killing vector $\xi^\mu$ on
	$\zTi_{d+1}$
	\begin{align}
	\label{eq:conscurrent}
	j^{\mu} = T\indices{^{\mu}_{\nu}}\xi^{\nu}\,,
	\end{align}
	leads to a conserved current in the absence of sources
	\begin{align}
	\cd_{\mu}j^{\mu} = \cd_{\mu} T\indices{^{\mu}_{\nu}}\xi^{\nu} +  T\indices{^{\mu}_{\nu}}\cd_{\mu} \xi^{\nu} = 0\,.
	\end{align}
	As discussed in Section~\ref{sec:carroll-struc-Ti}, the isometries of
	$\zTi_{d+1}$ include supertranslations with
	$\xi_{S} = S(\x)\pd_{\tau}$ which in particular contain the Poincaré
	translations. For these supertranslations, we obtain
	\begin{align}
	\cd_{\mu}T\indices{^{\mu}_{\tau}} S + T\indices{^{a}_{\tau}} \pd_{a}S=0, 
	\end{align}
	which can be satisfied only if
	\begin{align}
	\cd_{\mu}T\indices{^{\mu}_{\tau}}=0\,, \qquad T\indices{^{a}_{\tau}}=0\,.
	\end{align}
	We recognise the second of these as the carrollian boost Ward
	identity~\eqref{eq:boost-WI}. It is straightforward to check that
	restricting to the Poincaré translations would lead to the same
	conclusion. The Killing vectors associated with boosts and rotations
	are of the form $\xi = \xi^{a}(\x) \pd_{a}$, which yields
	\begin{align}
	\cd_{\mu} T\indices{^{\mu}_{a}}\xi^{a} +  T\indices{^{a}_{b}}\cd_{a}^{\HH}\xi^{b} = 0\,,
	\end{align}
	where we defined $\nabla_a^{\HH}$ as the covariant derivative on
	$\HH^d$ with connection components $\Gamma^a_{bc}$. If $\xi^b$ is a
	Killing vector on the hyperboloid, i.e., a boost or a rotation, then
	only the antisymmetric part of $T\indices{^{a}_{b}}$ will remain. The
	above equation can then be written
	\begin{equation}
	\label{eq:7}
	\cd_{\mu} T\indices{^{\mu}_{a}}\xi^{a} +  T\indices{^{[ab]}}\frac{1}{2}(\partial_a\xi{_b}-\partial_a\xi_{b}) = 0\,,
	\end{equation}
	where we used the metric on the hyperboloid $\HH^d$ to raise and lower the spatial indices.
	Taking $\xi^a$ to be the generator of a rotation $\xi^a_{J_{bc}}=x_b\delta^{a}_c-x_c\delta^{a}_b$ allows us to write
	\begin{equation}
	T_{[ab]}=\frac{1}{2}(\cd_{\mu} T\indices{^{\mu}_{a}}x_b-\cd_{\mu} T\indices{^{\mu}_{b}}x_a)\,,
	\end{equation}
	which leads to
	\begin{equation}
	\cd_{\mu} T\indices{^{\mu}_{a}}\left(2\xi^a+x_b\partial^a\xi^b-x^b\partial_b\xi^a\right)=0\,,
	\end{equation}
	when plugged back into \eqref{eq:7}. In order for this to vanish for all Killing vectors, we have to demand
	\begin{equation}
	\cd_{\mu} T\indices{^{\mu}_{a}}=0 \qquad \Rightarrow \qquad T_{[ab]}=0\,.
	\end{equation}
	Taken together, we find that the carrollian EMT is of the form
	\begin{equation}
	T\indices{^\mu_\nu}=\begin{pmatrix}T\indices{^\tau_\tau}&0\\
	T\indices{^\tau_a}& T\indices{^{a}_b}
	\end{pmatrix}\,,
	\end{equation}
	with $T^{ab}$ symmetrical, and furthermore obeys 
	\begin{subequations}
		\label{eq:carrollianconservation}
		\begin{align}
		\nabla_\tau T\indices{^\tau_\tau}&=0\qquad \Leftrightarrow \qquad \partial_\tau T\indices{^\tau_\tau}=0\\
		\nabla_\mu T\indices{^\mu_a}&=0 \qquad \Leftrightarrow \qquad \partial_\tau T\indices{^\tau_a}+\cd_{b}^{\HH}T\indices{^b_a}=0\,,
		\end{align}
	\end{subequations}
	which we recognise as the conservation equation~\eqref{eq:diffeo-WI}.
	Note in particular that these conditions are independent of the choice
	of the undetermined part of the connection. Using the conserved current \eqref{eq:conscurrent} and the invariant
	top-form \eqref{eq:volumeform}, we can define conserved charges as
	integrals over the hyperboloid
	\begin{equation}
	\label{eq:consQ}
	Q[\xi]=\int_{\mathbb{H}^d} j \cdot \varepsilon=\int_{\mathbb{H}^d} \varepsilon^{(d)}\,j^\tau\,.
	\end{equation}
	In particular, we find for supertranslations
	\begin{equation}
	\label{eq:superTcharge}
	Q[S]=\int_{\mathbb{H}^d}\varepsilon^{(d)}\,T\indices{^\tau_\tau}S\,,
	\end{equation}
	and for boosts and rotations
	\begin{equation}
	\label{eq:rotcharge}
	Q[\xi^a]=\int_{\mathbb{H}^d} \varepsilon^{(d)}\, T\indices{^\tau_a}\xi^a\,,
	\end{equation}
	where the $\xi^a$ are the vectors $\xi_{B_a}$ and $\xi_{J_{ab}}$ given
	in \eqref{eq:fund-vector-fields}.

	\subsection{An example: the energy-momentum tensor of massive scalars}
	\label{sec:EMT-scalar}
	
	As an illustration of the above, we will consider the carrollian EMT
	associated with massive real scalars. We obtain it in two ways, first
	by pushing the corresponding relativistic EMT from Minkowski space
	$\MM^{d+1}$ to $\zTi_{d+1}$, and then from a carrollian action
	principle.
	
	We thus start by considering a free massive scalar field in Minkowski space. The corresponding EMT is given by
	\begin{equation}
	\label{eq:SETMink}
	T_{\mu\nu}=\partial_\mu\phi\partial_\nu\phi-\frac{1}{2}g_{\mu\nu}\left(m^2\phi^2+\partial_\mu\phi\partial^\mu\phi\right)\,.
	\end{equation}
	We want to push this tensor to timelike infinity to obtain an example of a carrollian EMT. In hyperbolic coordinates we have
	\begin{subequations}
		\begin{align}
		\label{eq:SEThyp}
		T_{\tau\tau}&= \frac{\tau^{-d}}{2}\left((\partial_\tau\hat{\phi})^2+m^2\hat{\phi}^2-d\tau^{-1}\hat{\phi}\partial_\tau\hat{\phi}+\tau^{-2}\left(\frac{d^2}{4}\hat{\phi}^2+\partial^a\hat{\phi}\partial_a\hat{\phi}\right)\right)\,,\\
		T_{\tau a}&=\tau^{-d}(\partial_\tau\hat{\phi}\partial_a\hat{\phi}-\tau^{-1}\frac{d}{2}\hat{\phi}\partial_a\hat{\phi})\,,\\
		T_{ab}
		&=\tau^{-d}\left(\partial_a\hphi\partial_b\hphi-\frac{1}{2}h_{ab}(m^2\hphi^2+(\partial_\tau\hphi)^2)\right)\\
		&\quad -\frac{d}{2}h_{ab}\tau^{-d+1}\hphi\partial_\tau\hphi+\frac{\tau^{-d}}{2}\left(\frac{d^2}{4}\hphi^2h_{ab}+\partial_a\hphi\partial_b\hphi\right)\,,\nonumber
		\end{align}
	\end{subequations}
	where we used the rescaled field $\hphi=\tau^{\frac{d}{2}}\phi$
	defined in \eqref{rescaling}. At leading nontrivial order we have
	\begin{subequations}
		\label{eq:phistress}
		\begin{align}
		\bar T\indices{^\tau_\tau}&=-\frac{1}{2}\left( (\partial_\tau\bar{\phi})^2+m^2\bar{\phi}^2\right)=-m^2\phi^- \phi^+ + \mathrm{h.c.}\,,\\
		\bar T\indices{^\tau_a}&=-\partial_\tau \bar \phi\, \partial_a \bar \phi
		=im \left(\phi^+ \partial_a \phi^+\, e^{2im\tau}+\partial_a \phi^-\,\phi^+  \right)+ \mathrm{h.c.}\,,\\
		\bar T\indices{^a_b}&=\frac{1}{2} \delta^a_b\left((\partial_\tau\bar{\phi})^2-m^2\bar{\phi}^2\right)=-m^2 \delta^a_b (\phi^+)^2\, e^{2im\tau}+ \mathrm{h.c.}\,,
		\end{align}
	\end{subequations}
	where we used the solutions of the leading equations of
	motion~\eqref{induced carrollian field}, and where
	$\phi^- = (\phi^+)^*$ since we are dealing with real scalars. This
	carrollian EMT can be shown to satisfy the carrollian conservation
	equations \eqref{eq:carrollianconservation}.
	
	Although we started with the EMT of a free theory in
	\eqref{eq:SETMink}, the resulting carrollian EMT \eqref{eq:phistress}
	will not change if a potential is included. As discussed below
	\eqref{wave equation}, the leading order field in a large $\tau$
	expansion is still given by the free field. Furthermore, contributions
	coming from the potential will fall off faster than $\tau^{-d}$ so
	that the leading order terms in \eqref{eq:phistress} stay unaltered.
	
	As we shall now demonstrate, the above is nothing but the archetypal
	electric Carroll EMT. The action of an electric Carroll scalar field
	theory coupled to an arbitrary carrollian background is
	\begin{equation}
	\label{eq:scalarelectric}
	S_{\text{e}} = -\frac{1}{2}\int d^{d+1}x\,e\left[ -n^\mu n^\nu \D_\mu \bar\phi \D_\nu \bar\phi + m^2\bar\phi^2 \right]\,.
	\end{equation}
	Using the variational relations
	\begin{equation}
	\delta e = e\left( n^\mu \delta \tau_\mu + \frac{1}{2}\gamma^{\mu\nu}\delta q_{\mu\nu} \right)\,,\qquad \delta n^\mu =  - n^\mu n^\nu\delta\tau_\nu - \gamma^{\mu\nu}n^\rho \delta q_{\rho\nu}\,,
	\end{equation}
	we find that
	\begin{equation}
	\begin{split}
	T^\mu{_\nu} &= -\frac{1}{2}n^\mu\tau_\nu ( n^\rho n^\sigma \D_\rho \bar\phi\D_\sigma \phi + m^2 \bar\phi^2 ) + \frac{1}{2} \gamma^{\mu\lambda} q_{\lambda\nu} (n^\rho n^\sigma \D_\rho \bar\phi\D_\sigma\bar\phi - m^2\bar\phi^2) \\
	&\quad - n^\mu \gamma^{\rho\lambda} q_{\lambda\nu} n^\sigma\D_\rho\bar\phi\D_\sigma\bar\phi\,,
	\end{split}
	\end{equation}
	which on $\zTi_{d+1}$ becomes~\eqref{eq:phistress}. Furthermore, the equation of motion for $\phi$ is 
	\begin{equation}
	n^\mu n^\nu \D_\mu \D_\nu \bar\phi = -m^2\bar\phi\,,
	\end{equation}
	which reduces to the Casimir equation~\eqref{eq:reducibility}. 
	
	Using the explicit form of the EMT one can establish that the charges
	\eqref{eq:consQ} with the EMT \eqref{eq:phistress} form a
	representation of the symmetry algebra of $\zTi$. The commutator of
	the carrollian fields
	\begin{equation}
	[\phi^+(\y),\phi^-(\y')]=\frac{i}{2m}[\bar \phi(\tau,\y),\partial_\tau \bar \phi(\tau,\y')]=\frac{1}{2m}\sqrt{1+\y^2}\, \delta^{(d)}(\y-\y')=:\frac{1}{2m}\hat{\delta}^{(d)}(\y-\y'),
	\end{equation}
	follows from the canonical commutation relations
	\eqref{eq:poissonbracketsa} and the identification \eqref{phi=a}; the
	delta function $\hat{\delta}^{(d)}(\y-\y')$ is associated to the
	volume form on the hyperboloid in $y^a$ coordinates
	\eqref{eq:volumeform}. It follows then that
	\begin{subequations}
		\begin{align}
		[Q[S],Q[S']]&=0,\\
		[Q[S],Q[\xi^a]]&=-i Q[\xi^a\cd_{a}^{\HH}S]-i\int_{\HH} \varepsilon^{(d)}S\bar T\indices{^a_b}\cd_{a}^{\HH}\xi^b=-i Q[\xi^a\cd_{a}^{\HH}S],\\
		[Q[\xi^a_1],Q[\xi^b_2]]&=i Q[\xi^b_2\cd_{b}^{\HH}\xi^a_1-\xi^b_1\cd_{b}^{\HH}\xi^a_2] \, ,
		\end{align}
	\end{subequations}
	which is indeed a representation of the symmetry algebra on
	$\zTi$.\footnote{Here and in the following, we assume the prescription
		that the quantum EMT is obtained from the classical EMT
		\eqref{eq:phistress} by symmetrisation, e.g.,
		$\bar T\indices{^\tau_\tau}=-2m^2\phi^- \phi^+\rightarrow
		-m^2(\phi^- \phi^++\phi^+ \phi^-)$.}
	
	More generally, the canonical commutation relations imply the
	following algebra of the carrollian EMT
	\begin{subequations}
		\label{eq:setalgebra}
		\begin{align}
		[\bar T\indices{^\tau_\tau}(\y),\bar T\indices{^\tau_a}(\y')]&=i\left(\partial_a\bar T\indices{^\tau_\tau}(\y)-\bar T\indices{^\tau_\tau}(\y)\partial'_a+\bar T\indices{^b_a}(\y)\partial'_b\right)\,\hat{\delta}^{(d)}(\y-\y')\,,\\
		[\bar T\indices{^\tau_\tau}(\y),\bar T\indices{^a_b}(\y')]&=-i\partial_\tau \bar T\indices{^a_b}(\y')\,\hat{\delta}^{(d)}(\y-\y')\,,\\
		[\bar T\indices{^\tau_a}(\y),\bar T\indices{^\tau_b}(\y')]&=i\left(-\bar T\indices{^\tau_a}(\y)\partial'_b+\bar T\indices{^\tau_b}(\y')\partial_a+\partial_a\bar T\indices{^\tau_b}(\y)-\partial_b\bar T\indices{^\tau_a}(\y)\right)\hat{\delta}^{(d)}(\y-\y')\,,\\
		[\bar T\indices{^\tau_a}(\y),\bar T\indices{^b_c}(\y')]&=i\left(\partial_aT\indices{^\tau_\tau}(\y)\delta^{b}_c-\bar T\indices{^\tau_\tau}(\y')\delta^{b}_c\partial_a+\bar T\indices{^b_c}(\y')\partial_a\right)\hat{\delta}^{(d)}(\y-\y')\,,\\
		[\bar T\indices{^\tau_\tau}(\y),\bar T\indices{^\tau_\tau}(\y')]&=[\bar T\indices{^a_b}(\y),\bar T\indices{^c_d}(\y')]=0\,. 
		\end{align}
	\end{subequations}
	This algebra is arguably quite different from the EMT algebra of a
	lorentzian QFT as determined originally in
	\cite{Schwinger:1962wd,Schwinger:1963zza,Deser:1967zzf}. In
	particular, the vanishing of the
	$[\bar T\indices{^\tau_\tau}(\y),\bar T\indices{^\tau_\tau}(\y')]$
	commutator can be regarded as hallmark of carrollian symmetry
	\cite{Henneaux:2021yzg}. In a lorentzian field theory, the EMT algebra
	is universal, except for the commutator of purely spatial components,
	and can be derived by coupling the theory to a curved background and
	subsequent variation of the metric. It would be interesting to
	similarly analyse the universality of \eqref{eq:setalgebra} and
	investigate the possibility of central terms by applying an analogous
	procedure using the background carrollian structure.
	
	\subsection{Relation to BMS charges and soft graviton theorems}
	\label{sec:soft-graviton}
	
	In previous subsections we have shown that conserved charges are
	associated to Killing symmetries of $\zTi_{d+1}$. In four spacetime
	dimensions ($d=3$) it turns out that BMS symmetries which consist of
	supertranslations \cite{Bondi:1962px,Sachs:1962zza,Sachs:1962wk} and
	Lorentz transformations can be identified as a subset of these
	symmetries of $\zTi_4$. In fact the modern view is that a single BMS
	symmetry group acts simultaneously on all asymptotic regions of
	asymptotically flat spacetimes
	\cite{Campiglia:2015kxa,Campiglia:2015lxa,Troessaert:2017jcm,Prabhu:2019fsp,Prabhu:2021cgk,Capone:2022gme,Compere:2023qoa},
	with a single function $f(\Omega)$ valued over the celestial sphere
	$\mathbb{CS}^2$ to parametrise supertranslations. In particular
	supertranslations form a subset of the $\zTi$-supertranslations
	$S(\x)$ which are determined from $f(\Omega)$ through the formula
	\cite{Campiglia:2015kxa,Campiglia:2015lxa}
	\begin{equation}
	\label{BMS supertranslation}
	S(\x)=\int_{\mathbb{CS}^2} d\Omega\, G(\x,\Omega) f(\Omega)\,,
	\end{equation}
	where $G(\x,\Omega)$ is a `boundary-to-bulk' propagator satisfying the gauge-fixing equation
	\begin{equation}
	\left( \triangle_{\HH}-3 \right)\, G(\x)=0\,, 
	\end{equation}
	and chosen in such a way that $f(\Omega)$ is the asymptotic boundary
	value of $S(\x)$ along the asymptotic boundary
	$\partial \HH^3\cong \mathbb{CS}^2$. Hence with this particular choice
	of $S(\x)$ the corresponding charges \eqref{eq:superTcharge} generate
	the BMS supertranslations of the massive carrollian fields (i.e.,
	massive particles), while the charges \eqref{eq:rotcharge} generate
	their Lorentz transformations.
	
	It is by now a well-known result that Weinberg's soft graviton theorem
	\cite{Weinberg:1965nx} is the Ward identity associated with
	supertranslation symmetry
	\cite{Strominger:2013jfa,He:2014laa,Campiglia:2015kxa}. We can sketch
	the proof of this result in the following way.\footnote{We invite the
		reader to consult \cite{Donnay:2022hkf,Agrawal:2023zea} for a recent
		detailed discussion and for further references.} Working with a
	theory of gravity and interacting matter whose phase space is such
	that there exist truly conserved BMS charges $Q(i^0)$ at spatial
	infinity, the latter can be re-expressed as a sum of a BMS fluxes
	going through $\scri^\pm$ and a remaining BMS charge at $i^\pm$,
	\begin{equation}
	Q(i^0)=Q(i^+)-F(\scri^+)=Q(i^-)+F(\scri^-)\,.
	\end{equation}
	Invariance of the $\mathcal{S}$-matrix under BMS symmetries,
	\begin{equation}
	[Q(i^0)\,,\mathcal{S} ]=0\,,
	\end{equation}
	then yields
	\begin{equation}
	\label{Ward identity}
	\langle \text{out}| (F(\scri^+)-Q(i^+) )\, \mathcal{S}+\mathcal{S}\, ( Q(i^-)+F(\scri^-) ) |\text{in}\rangle=0\,.
	\end{equation}
	The soft graviton theorem is then obtained by acting with
	$F(\scri^\pm)$ and $Q(i^\pm)$ on in and out states. More specifically
	each flux can be decomposed into so-called \textit{hard} and
	\textit{soft} contributions
	$F(\scri^\pm)=F^{\text{hard}}(\scri^\pm)+F^{\text{soft}}(\scri^\pm)$.
	The action of the soft flux $F^{\text{soft}}(\scri^+)$ is to
	explicitly add a soft graviton to the outgoing states, while the
	action of $F^{\text{hard}}(\scri^+)$ and $Q(i^+)$ generate the BMS
	transformations of the massless and massive external particle states,
	respectively. The same holds for the operators acting on the in
	states. Writing this explicitly, manifestly reproduces Weinberg's soft
	graviton theorem
	\cite{Strominger:2013jfa,He:2014laa,Campiglia:2015kxa}.
	
	Having laid down these facts it should not come as a surprise that the
	BMS supertranslation charges $Q(i^\pm)$ entering the (re-)derivation
	of Weinberg soft graviton theorem in the presence of massive scalar
	particles \cite{Campiglia:2015kxa} are nothing but the
	$\zTi$-supertanslation charges \eqref{eq:superTcharge} with the
	particular restriction \eqref{BMS supertranslation} on the symmetry
	parameter $S(\bm{x})$. Indeed upon using the carrollian EMT for
	massive carrollian scalar fields given in \eqref{eq:phistress}
	together with the relation \eqref{phi=a} between carrollian fields and
	particle operators, the charge \eqref{eq:superTcharge} takes the
	explicit form
	\begin{equation}
	\label{QS}
	Q[S]=\int_{\HH^3} \varepsilon\, \bar T\indices{^\tau_\tau} S=-2m^2 \int_{\HH^3} \varepsilon\, \phi^- \phi^+\, S  =-\frac{m^3}{2(2\pi)^3} \int_{\HH^3} \varepsilon\, a^\dagger a\, S \,, 
	\end{equation}
	which exactly agrees with Eq.~(36) in \cite{Campiglia:2015kxa} up to a
	choice of normalisation. This demonstrates that $\zTi_4$ provides the
	natural asymptotic structure for massive fields even when it comes to
	fully interacting theories. This discussion should hold for massive
	spinning particles as well although this has not been explicitly
	studied in the literature.
	
	\section{Discussion and outlook}
	\label{sec:outlook}
	
	Motivated by flat space holography we have provided a carrollian
	description of massive particles. We have done this from an intrinsic
	perspective, as in Section~\ref{sec:ads-carr-intr}, and as a late time
	limit of Minkowski space, as in Section~\ref{sec:relat-fields-mink}.
	This provides an extrapolate dictionary for (symmetric) massive
	integer spin $s$ fields in generic dimension. We have also related
	this carrollian description to the plane wave, momentum and celestial perspectives,
	cf.~Figure~\ref{fig:triangle}. This work also shows that there are
	interesting applications for Carroll physics beyond the more
	extensively studied flat case. In addition, we have provided an intrinsic
	definition of the carrollian energy-momentum tensor on $\zTi_{d+1}$
	and discussed the associated $\zTi$-supertranslation charges along with their relation to soft
	gravitation theorems.
	
	This work opens various interesting avenues for further exploration:
	\begin{description}
		\item[Massive scattering amplitudes] A natural extension of the
		present work is the study of correlation functions of massive
		carrollian fields on $\zTi$, and their relation to the standard
		scattering amplitudes of massive particles. As a simple example we
		can consider the free 1-1 scattering of a massive scalar particle,
		with $\mathcal{S}$-matrix element simply given by the Lorentz invariant inner
		product
		\begin{equation}
		\langle p_1 |\mathcal{S}|p_2 \rangle=\sqrt{m^2+\| \bm{p}_1\|^2}\, \delta^{(d)}(\bm{p}_1-\bm{p}_2)= \left(\frac{m \sinh r_1}{r_1} \right)^{1-d}\, \delta^{(d)}(\x_1-\x_2)\,,
		\end{equation}
		where the last equality follows from the natural parameterisation of
		the massive momenta given in \eqref{eq:p-eigen}. This simple
		amplitude can be viewed as a two-point function for the
		corresponding carrollian field
		\begin{equation}
		\langle \phi(\tau_1,\x_1) \phi(\tau_2,\x_2) \rangle=\mathcal{N}\, \left(\frac{ \sinh r_1}{r_1} \right)^{1-d}\, \delta^{(d)}(\x_1-\x_2)\,,
		\end{equation}
		with $\mathcal{N}$ some normalisation factor, since for any of the
		Poincaré transformations $\delta \phi$ given in
		\eqref{eq:action-on-fields} it can be shown to satisfy the
		integrated Ward identity
		\begin{equation}
		\langle \delta \phi(\tau_1,\x_1) \phi(\tau_2,\x_2) \rangle + \langle \phi(\tau_1,\x_1) \delta \phi(\tau_2,\x_2) \rangle=0\,.
		\end{equation}
		This correspondence between massive scattering amplitudes and
		carrollian correlation functions on $\zTi$ should hold generally as
		it is a simple consequence of the covariance of the
		$\mathcal{S}$-matrix together with the direct relation between
		massive particle states and carrollian fields discussed in this
		work. Note that scattering amplitudes are defined on the momentum
		mass-shell $\HH^d$, and are therefore the local observables of a
		codimension-one theory ``dual'' to field theory on
		$\mathbb{M}_{d+1}$. Furthermore the corresponding carrollian
		correlators are simple uplifts of the amplitudes from $\HH^d$ to
		$\zTi_{d+1}$. Nevertheless it will be of high interest to study
		whether this new perspective yields new ways to constrain and
		construct massive scattering amplitudes. We point out that the
		analogous connection between massless scattering amplitudes and
		correlation functions of carrollian conformal fields at null
		infinity $\scri$ has been the subject of numerous recent
		works~\cite{Bagchi:2016bcd,Banerjee:2018gce,Banerjee:2019prz,Donnay:2022aba,Bagchi:2022emh,Donnay:2022wvx,Bagchi:2023fbj,Saha:2023hsl,Salzer:2023jqv,Nguyen:2023vfz,Saha:2023abr,Nguyen:2023miw,Bagchi:2023cen,Mason:2023mti,Chen:2023naw}.
		
		\item[General scattering] More generally, this work also opens the
		possibility to study general scattering between massive and massless
		particles from a carrollian perspective. The massive and massless
		carrollian fields seem to live on different asymptotic boundaries,
		but a careful analysis could make it possibly to match them, see,
		e.g.,~\cite{Compere:2023qoa} and references therein for work in this
		direction. Another possibility would be to use the embedding space
		picture we discuss in Section~\ref{sec:PoinTiem} where spatial,
		timelike and null infinity embed in a larger connected space.
		
		\item[Superrotations and subleading soft graviton theorems] The
		question naturally arises whether $\zTi_4$ also captures the
		extended version of the BMS group that includes Virasoro
		superrotations in addition to supertranslations
		\cite{Barnich:2009se,Barnich:2010eb}. Similarly the subleading soft
		graviton theorem \cite{Cachazo:2014fwa} and associated loop
		corrections \cite{Bern:2014oka,Sahoo:2018lxl} are also known to
		derive from the Ward identities of superrotation symmetries
		\cite{Adamo:2014yya,Kapec:2014opa,Campiglia:2015kxa,He:2017fsb,Donnay:2022hkf,Agrawal:2023zea}.
		However the group of carrollian isometries of $\zTi_4$ does not
		contain symmetries that could be identified with Virasoro
		superrotations (beyond the standard Lorentz
		transformations, cf.,~also our discussion at the end of
		Section~\ref{sec:carroll-struc-Ti}). To better understand the
		origin of Virasoro superrotations it is useful to project from
		$\zTi_4$ down to its hyperbolic base $\HH^3$ also known as euclidean
		$\zAdS_3$. In that case it is known that Virasoro symmetries arise
		as asymptotic symmetries of three-dimensional gravity with $\zAdS_3$
		asymptotics \cite{Brown:1986nw}. Although the latter are conformal
		isometries of the celestial sphere
		$\partial \HH^3 \cong \mathbb{CS}^2$, away from this asymptotic
		boundary they do not leave the metric on $\HH^3$ invariant. This is
		no issue however since the metric in the bulk of $\HH^3$ is a
		dynamical gravitational field allowed to fluctuate. This suggests
		that Virasoro superrotations arise at $\zTi_4$ when the metric
		\eqref{eq:invariants} is allowed to fluctuate away from
		$\partial \HH^3\cong \mathbb{CS}^2$, which would appear natural in a
		theory of gravity with flat asymptotics. In fact the only kind of
		fluctuations needed to witness the appearance of Virasoro
		superrotations are locally pure diffeomorphisms. Indeed an
		implementation of the above ideas has been presented in
		\cite{Ball:2019atb}. In short the description of superrotation
		symmetries and charges requires to go beyond isometries of $\zTi_4$
		by allowing for metric fluctuations, and therefore goes beyond the
		kinematic framework presented here.
		
		To go beyond the kinematical framework of this work one can
		``gauge'' the $\zTi_4$ symmetries, i.e., builds theories which allow
		for curved and dynamical geometries. This is the analogue of gauging
		Minkowski space (often referred to as ``gauging the Poincaré
		algebra'') which leads to Einstein gravity in the first order
		formalism. For $\zTi_4$ (or equivalently AdS Carroll) this was done
		in~\cite{Figueroa-OFarrill:2022mcy} (see
		also~\cite{Hartong:2015xda}) and leads to a carrollian theory of
		gravity.\footnote{It was shown in~\cite{Campoleoni:2022ebj} that
			this is equivalent to ``magnetic'' Carroll
			gravity~\cite{Bergshoeff:2017btm} with cosmological constant.} An
		asymptotic analysis of this theory~\cite{Perez:2022jpr} leads to the
		expected BMS$_{4}$ symmetries and also shows that it allows for
		superrotations. It would be interesting to see whether this gauged
		$\zTi_4$ theory naturally arises at timelike infinity if gravity in
		the bulk asymptotically flat spacetime is dynamical.
		
		\item[Detector operators] The integral along null infinity of
		(weighted) components of the EMT in asymptotically flat spacetime
		can be interpreted as \emph{detector operators} measuring properties
		of the outgoing flux of massless particles in a given
		state.\footnote{Having had a long history in the context of QCD
			(e.g.\ \cite{Collins:1981uw}), such non-local operators have more
			recently been studied in the context of CFTs (e.g.\
			\cite{Kravchuk:2018htv,Caron-Huot:2022eqs}) and gravity
			\cite{Gonzo:2020xza}.} Here we will argue that the carrollian EMT
		discussed in Section \ref{sec:EMT-and-conservation-laws} can be used
		to construct idealised detectors for massive particles. As a
		concrete example, consider the final state of a scattering
		experiment consisting of $n$ particles of identical mass $m$
		\begin{equation}
		\label{final state}
		\ket{n}=\ket{p_1,\ldots, p_n}=a^\dagger_{p_1}\ldots a^\dagger_{p_n}\ket{0}=\left(m^{1-d/2}2(2\pi)^{d/2}\right)^n\phi^{-}(\y_1)\ldots \phi^{-}(\y_n)\ket{0}. 
		\end{equation}
		From the normal-ordered carrollian energy density operator
		\begin{equation}
		:\bar T\indices{^\tau_\tau}(\y):=-2m^2\phi^-(\y)\phi^+(\y)\,,
		\end{equation}
		one can construct an operator that just measures the total mass of
		the final state
		\begin{equation}
		-\int_{\HH}\varepsilon^{(d)}:T\indices{^\tau_\tau}(\y):\ket{n}=n \,m \,\ket{n}\,,
		\end{equation}
		or the total energy
		\begin{equation}
		-\int_{\HH}\varepsilon^{(d)}\,\sqrt{1+\y^2}\,:T\indices{^\tau_\tau}(\y):\ket{n}=\sum^n_{i=1} \,m \sqrt{1+\y^2_i} \,\ket{n}=\sum^n_{i=1} E_i \,\ket{n}\,.
		\end{equation}
		These are of course nothing but the supertranslation charge for
		$S=-1$ and global time translation $S=-\cosh r=-\sqrt{1+\y^2}$.
		
		At particle colliders what is typically measured is the total energy
		density which leaves along a given spatial direction $\bm{n}$
		from the origin of the detector. This implies that the spatial
		momentum of a detected particle was proportional to $\bm{n}$,
		namely $\bm{p}=\|\bm{p}\| \bm{n}$. Now recall that the
		asymptotic leaf $\HH_{d}$ is identified with the momentum mass shell
		of the particles through the critical point equation \eqref{eq:18}.
		Using the natural parametrisation $p^\mu(\y)=m(\sqrt{1+\y^2},\y)$
		that follows from \eqref{eq:hyppar}--\eqref{k hyppar}, we
		equivalently have $\y=\|\y\| \bm{n}$ for any of the detected
		particles, and the corresponding total energy density left along the
		direction $\bm{n}$ is therefore obtained by integrating
		$:T\indices{^\tau_\tau}(\y):$ over the radial coordinate $\|\y\|$ on
		the hyperboloid,
		\begin{equation}
		\label{eq:energyflow}
		\mathcal{E}(\bm{n})\equiv-\int^{\infty}_0\dd \|\y\| \|\y\|^{d-1}:T\indices{^\tau_\tau}(\|\y\|\bm{n}):\,. 
		\end{equation}
		The application of this operator on the final state \eqref{final
			state} yields
		\begin{equation}
		\mathcal{E}(\bm{n})\ket{n}=\sum^{n}_{i=1}E_i\, \delta^{(d-1)}(\bm{n}-\bm{n}_i)\ket{n}\,,
		\end{equation}
		which as expected gives the sum of the energies of the particles
		whose spatial momenta were aligned to $\bm{n}$. The operator
		$\mathcal{E}(\bm{n})$ provides a massive analogue to the
		well-known energy-flow operator counting the total energy carried by
		massless particles along the direction $\bm{n}$; see, e.g.,
		\cite{Korchemsky:1999kt} for applications of this operator to
		$e^+e^-$ annihilation. Note that it is also the charge
		\eqref{eq:superTcharge} associated with a $\zTi$-supertranslation
		$S_{\bm{n}}(\y)=-\sqrt{1+\y^2}\,
		\delta^{(d-1)}(\bm{n}_{\y}-\bm{n})$. The hallmark equation
		of carrollian physics, i.e., the vanishing of the
		$[\bar T\indices{^\tau_\tau}(\y),\bar T\indices{^\tau_\tau}(\y')]$
		commutator, implies that the energy-flow operator commutes
		\begin{equation}
		[\mathcal{E}(\bm{n}),\mathcal{E}(\bm{n'})] =0\,,
		\end{equation}
		i.e., the energy flow in different directions can be measured
		independently from each other. Finally,
		$\bar T\indices{^\tau_\tau}(\y)$ itself measures the energy as a
		function of the spatial momentum.
		
		The above only presents one example of conceivable detector
		operators for massive particles, the design of which depends on the
		questions to be answered in a given thought experiment. Although
		arguably not as rich due to the absence of collinear and soft
		divergences, it should be clear that the carrollian EMT and its
		algebra can be viewed as providing the basic building blocks for
		detector operators for massive particles.
		
		\item[Curved space] Our discussion was based on a fixed Minkowski
		background and it would be interesting to generalise to curved
		backgrounds. In the following, let us present a simple argument that
		the results of Section~\ref{sec:connection-with-bulk} should carry
		over.
		
		Assuming that the Ashtekar--Hansen definition of asymptotic flatness
		at spatial infinity \cite{Ashtekar:1978zz} in four dimensions can be
		adapted to $\zTi$, we require the existence of an unphysical,
		conformally related metric
		$\tilde{g}_{\alpha\beta}=\Omega^2g_{\alpha\beta}$ that is continuous
		at time-like infinity. In contrast to null infinity, the conformal
		factor is required to obey the properties
		\begin{equation}
		\label{eq:AHconditions}
		\Omega|_{i^+}=0,\qquad \tilde{\nabla}_\alpha\Omega|_{i^+}=0, \qquad (\tilde{\nabla}_\alpha\tilde{\nabla}_\beta\Omega+2 \tilde g_{\alpha\beta})|_{i^+}=0\,.
		\end{equation}
		Similar to the relation between the Ashtekar--Hansen construction
		and hyperbolic coordinates near spatial infinity (see Appendix B in
		\cite{Prabhu:2019daz} for a comprehensive discussion) we relate the
		conformal factor to a time coordinate as $\Omega=\tau^{-2}$. We can
		use this to perform an asymptotic analysis of the wave equation for
		a scalar field in four dimension. Writing the wave equation in terms
		of the unphysical metric yields
		\begin{equation}
		0=(g^{\alpha\beta}\nabla_\alpha\nabla_\beta -m^2)\phi=(\Omega^2\tilde{g}^{\alpha\beta}\tilde{\nabla}_\alpha\tilde{\nabla}_\beta-2 \Omega \tilde{g}^{\alpha\beta}\tilde{\nabla}_\alpha\Omega\tilde{\nabla}_\beta-m^2)\phi.
		\end{equation}
		Replacing the inverse metric using the third equation of
		\eqref{eq:AHconditions} one finds that the Laplacian is projected
		onto $-\partial^2_\tau$ to leading order in $\Omega$. Consequently,
		one finds the asymptotic equation
		\begin{equation}
		(-\partial^2_\tau-m^2)\phi+\mathcal{O}(\Omega^{1/2})=0,
		\end{equation}
		so that the asymptotic solutions are still given by the $\zTi$
		fields \eqref{induced carrollian field} in agreement with the
		strictly flat case.
		
		Clearly, this analysis depends on a bona fide extension of the
		Ashtekar--Hansen analysis to time-like infinity by analytic
		continuation. A more thorough discussion of the general curved case
		would therefore require a similar analysis of appropriate boundary
		conditions at time-like infinity along the of
		e.g.,~\cite{Compere:2023qoa}. 
		
		We end by noting that the work \cite{Borthwick:2023lye} demonstrated
		that generic asymptotically flat solutions to Einstein gravity can
		be compactified so that they give rise to curved versions (in the
		Cartan sense) of $\zTi$ at asymptotically late times. It would be
		interesting to make closer contact with this analysis and similarly
		interpret the late-time behaviour of fields on a curved bulk
		spacetime in terms of fields on these curved analogues of $\zTi$.
	\end{description}
	
	\acknowledgments
	
	We are grateful to Tim Adamo for valuable discussions. We thank José Figueroa-O’Farrill for collaboration at early stages of
	the project, and we are grateful to him for useful discussions. KN
	thanks Peter West for collaboration on closely related topics. The
	work of EH is supported by Villum Foundation Experiment project
	00050317, ``Exploring the wonderland of Carrollian physics''. The work
	of KN and JS is supported by postdoctoral research fellowships of the
	F.R.S.-FNRS (Belgium). Part of this project was revised and presented during the
	Programme ``Carrollian Physics and Holography'' at the Erwin-Schrödinger
	International Institute for Mathematics and Physics in April 2024.
	
	\addcontentsline{toc}{section}{\refname}

	\bibliographystyle{utphys} %~/texmf/bibtex/bst/
	\bibliography{bibl} %~/texmf/bibtex/bib/
	
\end{document}